\documentclass[prb,twocolumn,superscriptaddress,showpacs,preprintnumbers,amsmath,amssymb,floatfix]{revtex4-1}
\usepackage{epsfig}
\usepackage{auto-pst-pdf}
\usepackage{graphics} 
\usepackage{color}
\usepackage{amsmath}
\usepackage[english]{babel}
\usepackage{amssymb}

\newcommand{\ket}[1]{|#1\rangle}
\newcommand{\beq}{\begin{equation}}
\newcommand{\eeq}{\end{equation}}

\newcommand{\sandwich}[3]{\langle#1|#2|#3\rangle}

\date{\today}
\begin{document}

\title{Transport through two interacting resonant levels connected by a Fermi sea}

\author{Elena Canovi}
\affiliation{Institut f\"ur Theoretische Physik III, Universit\"at Stuttgart, Pfaffenwaldring 57, 70550 Stuttgart, Germany}

\author{Alexander Moreno}
\affiliation{Institut f\"ur Theoretische Physik III, Universit\"at Stuttgart, Pfaffenwaldring 57, 70550 Stuttgart, Germany}

\author{Alejandro Muramatsu}
\affiliation{Institut f\"ur Theoretische Physik III, Universit\"at Stuttgart, Pfaffenwaldring 57, 70550 Stuttgart, Germany}

\begin{abstract}
We study transport at finite bias, i.e.\ beyond the linear regime, through two interacting resonant levels connected by a Fermi sea, 
by means of time-dependent density matrix renormalization group.
We first consider methodological issues, like the protocol that leads to a current-currying state and the characterization of the steady state. At finite sizes both the current and the occupations of the interacting levels oscillate as a function of time. We determine the amplitude and period of such oscillations as a function of bias. We find that the occupations on the two dots oscillate with a relative phase which depends on the distance between the impurities and on the Fermi momentum of the Fermi sea, as expected for RKKY interactions.  Also the approximant to the steady-state current  displays oscillations as a function of the distance  between the impurities. Such a behavior can be explained by resonances in the free case. We then discuss the incidence of interaction on such a behavior. We conclude by showing the effect of the bias on the current, making connection with the one-impurity case.
\end{abstract}


  
 \maketitle
 \section{Introduction}
The study of quantum transport across nanostructures has been the subject of intense theoretical and experimental attention for decades. One of the most intensively studied systems is that of quantum dots, both because of their great experimental versatility and because they unveil an extremely rich physics,
as exemplified by the Kondo effect \cite{Hewson} in quantum dots
\cite{GoldhaberGordon_NAT98, vanderWiel_SCI00}. When considering a system of two quantum dots, a further interesting phenomenon emerges, the Rudermann-Kittel-Kasuya-Yosida  (RKKY) interaction~\cite{Ruderman_PR54,*Kasuya_PTP56,*Yosida_PR57,*VanVleck_RMP62,*Kittel}. It describes the indirect interaction between two magnetic impurities mediated by the electrons of the surrounding Fermi sea, and is characterized by oscillations related to the Fermi wavevector. The competition between the
RKKY interaction and the Kondo effect was studied in the frame of numerical renormalization group \cite{Jones_PRL87,*Jones_PRL88, *Jones_PRB89}, and conformal field theory \cite{affleck92,affleck95}. An experimental realization 
with two quantum dots coupled by a Fermi sea was meanwhile reported \cite{Craig_Sc04}. 

Recently, a great deal of progress was achieved towards the theoretical determination of steady-state transport properties focusing on a quantum dot described by the interacting resonant level model (IRLM) \cite{Boulat_PRL08,Karrasch_PRB10,Karrasch_EPL10,Kennes_PRB12,Kennes_pp12a,Kennes_PRB12b,Andergassen_PRB11,Schmitteckert_PRB04,Branschadel_AP10, Einhellinger_PRB12}, that consists of spinless fermions with a nearest-neighbor repulsive interaction for the sites adjacent to the dot. This model was studied with several theoretical techniques, ranging from integrable field theories and Bethe Ansatz (see Boulat  et al.~\cite{Boulat_PRL08}  and references therein),
functional renormalization group \cite{Karrasch_PRB10,Karrasch_EPL10,Kennes_PRB12,Kennes_pp12a,Kennes_PRB12b}, real-time renormalization group~\cite{Andergassen_PRB11}, to density-matrix renormalization group (DMRG) techniques~\cite{Schmitteckert_PRB04, Boulat_PRL08, Branschadel_AP10, Einhellinger_PRB12}. These works provide the I-V characteristics out of equilibrium at finite bias and up to large values of the interaction~\cite{Boulat_PRL08}, and a detailed knowledge of the relaxation dynamics \cite{Karrasch_PRB10,Karrasch_EPL10,Andergassen_PRB11,Kennes_PRB12,Kennes_PRB12b} in the regime of small interaction, including also the incidence of finite temperatures \cite{Kennes_pp12a}. The shot noise and the full counting statistics have been studied by means of exact diagonalization~\cite{Branschaedel_PRB10} (in the free case), DMRG and thermodynamical Bethe Ansatz \cite{Branschaedel_PRL10,Carr_PRL11}. Such an attention on a model that arguably cannot be experimentally realized in an electronic system is due to the fact that, in contrast to the
Anderson impurity model, the important energy scales of the problem are accessible and controllable in numerical simulations,
avoiding to deal with the Kondo scale, that requires high resolution in energy.
  
In contrast to the great attention devoted to the one impurity case,  little is known about the case with more impurities \cite{Schneider_pp06,Enss_PRB05,Costamagna_PRB08, Molina_EPJB05,Weinmann_EPJB08}. In particular, to the best of our knowledge, the case of two IRLs separated by a Fermi sea under a finite bias awaits still a theoretical treatment. Here we consider two leads modeled as tight-binding chains with uniform hopping,
coupled to two quantum dots interacting with their nearest-neighboring sites and a Fermi sea in between, focusing on the dynamics of the system when it is taken out of equilibrium with the application of a finite bias. The set-up is that of a 
quantum quench, where the initial state corresponds to the ground state of a Hamiltonian, but the time evolution is governed by a different (time independent) one. We considered two different protocols, where the bias is included either in the initial or in the final Hamiltonian. We discuss also the characterization of the steady-state and the incidence of finite-size effects.

We performed our studies by means of a time-dependent DMRG (t-DMRG) simulation \cite{White_PRL04,Daley_JSTAT04,schmitteckert04,
Schollwoeck_AP11}. This method allows to study the time evolution of the system up to intermediate times ($\sim 40\hbar/t_{0}$, where $t_{0}$ is the nearest-neighbor hopping between the sites of the leads) in a nonperturbative way. 
The time evolution of the current on each link of the chain and of the particle-density on the dots exhibits oscillations  whose frequency depends on the applied bias, as in the single dot case.  
In the present case   
the occupations on the two dots oscillate with a relative phase which depends on the distance between the impurities, both in the free and in the interacting case. This can be explained in terms of the RKKY interaction. The currents through 
the sites connecting the quantum dots to the leads show also 
oscillations but with a phase-shift with respect to the density. 
 These oscillations are a finite-size effect, as already discussed in the single dot case
\cite{Branschadel_AP10}, and vanish in the limit of infintiely long leads, as shown below.
For the approximant to the steady-state current we find that it oscillates as a function of the distance between the impurities. In the free case the behavior of the current can be understood in terms of resonances that appear in the transmission coefficient of a single particle propagating through the system. 
We then show the effect of interaction.
While it suppresses the resonances  found in 
the free case,
for strong interaction we find that
large oscillations of the current as a function of the interimpurity distance arise. 
 Finally we consider 
the I-V characteristics in the presence of two impurities, showing  also in this context the presence of negative conductance.
\par
This paper is organized as follows. Section \ref{sec:gen} is devoted to the discussion of methodological issues. In particular in Sec.\ \ref{ssec:model} we define the model, the observables and the numerical technique. We show the effect of different quench schemes and motivate our choice in Sec.\ \ref{ssec:qs}.
In Sec.\ \ref{ssec:timeav} we detail how the approximant of the steady-state current is obtained and benchmark our DMRG results for the one impurity system with those of  Boulat {\em et  al}.\ \cite{Boulat_PRL08}. Section \ref{sec:results} displays our results. 
In Sec.\ \ref{ssec:phaserelations}  the time evolution of the occupations and the currents is shown and its relation with RKKY interaction is discussed. 
In Sec.\ \ref{ssec:Rdependence} we concentrate on the approximant to the steady-state values of the current as a function of the distance. We consider first the free case, for which we establish a connection with the problem of transmission of a single particle propagating in the system, and then move to the interacting case. 
 In Sec.\ \ref{ssec:Negative} we discuss the I-V characteristics in the presence of interaction, comparing it with the one-impurity case~\cite{Boulat_PRL08}.
In Sec.~\ref{sec:conclusions} we summarize our results.
%
\section{Models and methods}\label{sec:gen}
\subsection{Hamiltonian and observables}\label{ssec:model}
\begin{figure}[!htb]
  \begin{center}
    \includegraphics[width=1.0\columnwidth]{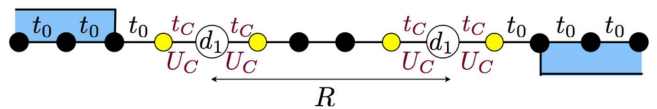}
    \caption{(Color online) Picture of the model Eq.~\ref{eq:hchain} for a system of $L=14$ sites and $R=5$. The shaded light blue areas indicate the presence of the bias (Eq.~\ref{eq:hbias}).}
    \label{fig:system}
  \end{center}
\end{figure}

We study a system characterized by the presence of two quantum dots at positions $d_{1}$ and $d_{2}$ separated by a 
distance $R\equiv d_{2}-d_{1}$. The region inbetween harbours a Fermi sea. The Hamiltonian of the whole system is given by
\beq\label{eq:hchain}
{\hat H}_{\rm chain}\equiv\hat H_{\rm D}+\hat H_{\rm T}+\hat H_{\rm F}\;,
\eeq
where
\begin{eqnarray}
\hat H_{\rm D}&=&-t_{{\rm C}}(\hat c^{\dagger}_{d_{1}-1}\hat c_{d_{1}}+\hat c^{\dagger}_{d_{2}}\hat c_{d_{2}+1}+{\rm H.c.})\nonumber\\
&&-t_{{\rm C}}(\hat c^{\dagger}_{d_{1}}\hat c_{d_{1}+1}+\hat c^{\dagger}_{d_{2}-1}\hat c_{d_{2}}+{\rm H.c.})\nonumber\\
&&+U_{\rm C}\sum_{\alpha=d_{1},d_{2}}\sum_{r=\pm 1}\left(\hat n_{\alpha}-\frac{1}{2}\right)\left(\hat n_{\alpha+r}-\frac{1}{2}\right)\;,\label{eq:hdots}
\end{eqnarray}
corresponds to the dots and their nearest-neighbors, where the interaction is present. The leads connecting to the quantum dot
are described by the tight-binding Hamiltonian $H_{\rm T}$,
\beq\label{eq:hleads}
\hat H_{\rm T}=-t_{0}\sum_{j=1}^{d_{1}-2}\hat c^{\dagger}_{j}\hat c_{j+1}-t_{0}\sum_{j=d_{2}+1}^{L-1}\hat c^{\dagger}_{j}\hat c_{j+1}+{\rm H. c.}
\; .
\eeq
Furthermore, the Fermi sea is described by the Hamiltonian $H_{\rm F}$,
\beq\label{eq:cregion}
\hat{H}_{\rm F}=-t_{0}\sum_{j=d_{1}+1}^{d_{2}-2}(\hat c^{\dagger}_{j}\hat c_{j+1}+{\rm H. c.}) 
\eeq
In what follows we call the sites at positions $c_1 \equiv d_{1}-1$ and $c_2 \equiv d_{2}+1$ contacts.
The total number of sites of the system is given by $L$, which we take even. If $R$ is odd, we choose the position of the dots such that the left  and the right leads have the same number of sites. 
If $R$ is even the position of the dots is given by $(L-R)/2+1$ and $(L+R)/2+1$, implying that the left lead has one more site with respect to the right one.
In Eqs.\ (\ref{eq:hdots}) - (\ref{eq:cregion})
we have $\hat n_{j}=\hat c^{\dagger}_{j}\hat c_{j}$,
where $\hat c^{\dagger}_{j}$ ($\hat c_{j}$) are creation (annihilation) operators for spinless fermions,
$U_{{\rm C}}$ is the interaction coupling the dots and their nearest neighbors, $t_{\rm C}$ is the hopping between the dot and its nearest-neighbors. The hopping elements in the leads and in the Fermi sea are all set to $t_{0}$. Energies are measured in units of $t_{0}$ and time in units of $\hbar/t_{0}$. 
The number of particles in the system is $N$ and we define the average density
of particles as $\rho\equiv N/L$. When not explicitly specified, we assume the system at half-filling.  
We also define $L_{\rm c}\equiv R-1$, $N_{\rm c}$ and $\rho_{\rm c}\equiv N_{\rm c}/L_{\rm c}$  as the number of sites, the number of particles and the density in the central region (from site $d_{1}+1$ to $d_{2}-1$), respectively.
The system is depicted in Fig.~\ref{fig:system}.
\par

As it will be discussed in more detail in Sec.\ \ref{ssec:qs},
we will follow the transport process in the frame of a quantum quench, where a given initial state $\ket{\Psi_{0}}$ evolves in time under the action of a given Hamiltonian, such that the state of the system at a time $\tau$ is $\ket{\Psi(\tau)}=\exp(-i\hat H\tau)\ket{\Psi_{0}}$. Accordingly, the time-dependent occupations on each site are given by 
\beq
n_{j}(\tau)\equiv \sandwich{\Psi(\tau)}{\hat n_{j}}{\Psi(\tau)}\;.
\eeq
The current on each bond connecting nearest-neighbor sites can be
obtained as:
\beq
I_{j} = i \frac{e}{\hbar} t_{j} \langle \Psi(\tau) |
(
\hat c^{\dagger}_{j}\hat c_{j+1}-\hat c^{\dagger}_{j+1}\hat c_{j}
) | \Psi(\tau) \rangle
\label{eq:current}\;,
\eeq
where $e$ is the electron charge, $t_{j}$ is the hopping on the bond connecting sites $j$ and  $j+1$.
\par
The results presented in this work are obtained with 
t-DMRG~\cite{White_PRL04,Daley_JSTAT04,schmitteckert04,Schollwoeck_AP11}.
We typically simulate systems with $L\sim100$ sites. In order to implement the time evolution, 
we use the Trotter decomposition~\cite{Daley_JSTAT04,White_PRL04,AH_PRB06}. 
Our code is adaptive~\cite{Daley_JSTAT04,White_PRL04,AH_PRB06}, meaning that the number of states used at each time step changes dynamically keeping the discarded weight
below a given threshold. The maximum number of states  used  in our computation is $m\sim1000$ and the discarded weight $\varepsilon$ is kept below $\sim 10^{-7}$. In the absence of interactions we employ also exact diagonalization.  Comparing the latter and DMRG for typical values of $m$ and $L$ we find that the relative error of the occupations is less than $10^{-4}$ for times  $\lesssim 40 \hbar /t_0$, while for the currents it is always less
than $10^{-3}$ in the same interval of time.

\subsection{Quench schemes}\label{ssec:qs}
\begin{figure}[!htb]
  \begin{center}
    \includegraphics[width=1.0\columnwidth]{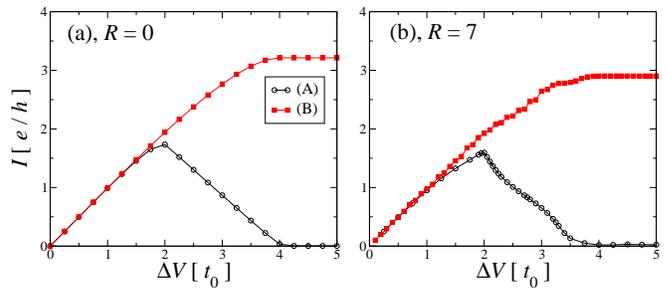}
    \caption{(Color online) I-V characteristics for a system with one (a) and two impurities (b). Black empty circles and red full squares refer to quench schemes (A)  and 
    (B) respectively. All the curves are obtained with $L=100$, except for $R=7$, scheme (A), for which $L=300$ sites are used. Data for $t_{\rm C}=0.8 t_{0}$ and $U_{\rm C}=0.0$. 
    The current is shown in absolute value.}
    \label{fig:IVR7}
  \end{center}
\end{figure}

In order to initiate transport processes in the system described by Eq.\ (\ref{eq:hchain}), a bias $\Delta V$ has to be applied on
the left and the right lead. It is described by:
\beq\label{eq:hbias}
\hat H_{\rm B}= \frac{\Delta V}{2}\left(\sum_{j=1}^{d_{1}-1}\hat n_{j}-\sum_{j=d_{2}+1}^{L}\hat n_{j}\right)\;.
\eeq
As previously discussed for a single impurity \cite{Branschadel_AP10}, we can start with the ground state of 
${\hat H}_{\mbox{\footnotesize chain}}$ and follow the evolution of the system dictated by a Hamiltonian ${\hat H}_{\mbox{\footnotesize chain}} + H_{\rm B}$. We denote such a procedure scheme (A). In such a scheme, however, the bounded nature of the spectrum of a lattice model becomes evident whenever the bias exceeds the bandwidth. In that case, there are no states available for transport through the system, as shown in Fig.\ \ref{fig:IVR7} (the determination of the currents depicted will be discussed in detail in Sec.\ \ref{ssec:timeav}).
It was suggested previously \cite{Boulat_PRL08,Branschadel_AP10}, that in order to avoid such an artifact of a lattice model, the opposite scheme can be used, namely, the initial state is the ground state of ${\hat H}_{\mbox{\footnotesize chain}} + H_{\rm B}$, and the evolution is studied switching off $H_{\rm B}$. As shown in Fig.\ \ref{fig:IVR7}, such a quench scheme leads to a saturation of the attained current, with similar behavior for a single impurity or two of them with a Fermi sea inbetween. 
The current in scheme (B) saturates at large values of the bias, because of the finite bandwidth of the system~\cite{Cini_PRB80,Branschadel_AP10}. 
For
$\Delta V$ smaller than half the bandwidth, both schemes lead to the same result. Moreover, for the whole range of biases studied in the one-impurity case \cite{Boulat_PRL08}, the I-V curves can be brought in this way to coincide with analytical results from conformal field theory.

In scheme (B) the initial state is characterized by a particle imbalance between the left and right lead, due to the presence of the bias, and the distribution of particles in the central region is not uniform. However, we find $\rho_{\rm c}=\rho$  if the system is at half filling. In the other cases there is a discrepancy which can be controlled by performing a finite-size scaling.

In the rest of the work we choose quench scheme (B) because 
it avoids the artifact introduced by a bounded spectrum.
\subsection{Time averages}\label{ssec:timeav}
As already discussed in the Refs.\ \onlinecite{Branschadel_AP10} and \onlinecite{Nuss_pp13} in the case of a single quantum dot, 
the time evolution of a current in a finite system is affected in various ways. On the one hand, right after switching the bias on (or off), there is a transient time, where the current grows from zero to a quasi-stationary state. On the other hand, at long times, the current bounces back at the ends of the system.
\begin{figure}[!htb]
  \begin{center}
    \includegraphics[width=0.9\columnwidth]{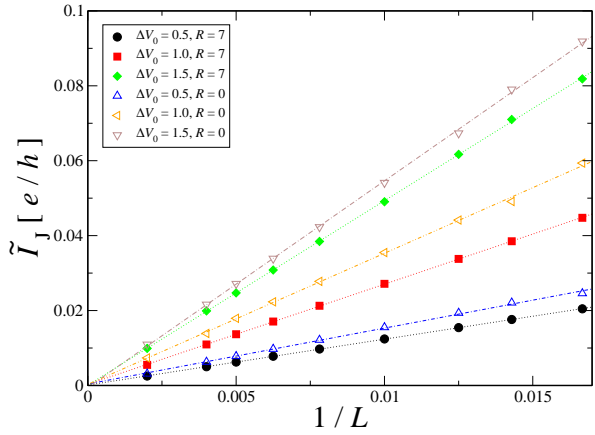}
    \caption{(Color online) Finite size scaling of the oscillation amplitudes $\tilde I_{\rm J}$ from cosine fits as discussed in main text, extracted from the left-contact current $I_{c1}$. Data refer to a system with $t_{\rm C}=0.8 t_{0}$, $U_{\rm C}=0$, $R=0$ (empty symbols) and $R=7$ (full symbols).}
    \label{fig:jos2}
  \end{center}
\end{figure}
In the intermediate quasi-stationary state, periodic variations previously denoted Josephson oscillations \cite{Branschadel_AP10}, due to their similarity with the ones in a Josephson junction, appear with a period $T_{\rm J}	\equiv 1/\nu_{\rm J}=2\pi/\Delta V$ determined by the bias, with an amplitude that vanishes \cite{Branschadel_AP10} as $1/L$.
Hence, in the free case one can extract an approximant to the steady-state current fitting the Josephson oscillations with a cosine function~\cite{Branschadel_AP10} of the form $I_{\alpha}(\tau)=\tilde I+\tilde I_{\rm J}\cos(2\pi\tau/T_{\rm J}+\tilde\varphi)$,
where $\alpha=c_1$ or $c_2$ denotes the left or right contact, and the free parameters of the fit are $\tilde I$, $\tilde I_{\rm J}$ and $\tilde\varphi$. 
\par
In the case of two impurities without interaction we find the same time scales, with minor differences. In particular the transient time also depends on the distance between 
the two impurities, and the amplitude of the Josephson oscillations is also affected by $R$. Nevertheless, as we show in Fig.~\ref{fig:jos2}, it is still possible to extract the approximant to the steady-state current by fitting the Josephson oscillations as mentioned above, obtaining an amplitude that also vanishes in the thermodynamic limit.  
\begin{figure}[!htb]
  \begin{center}
    \includegraphics[width=1.0\columnwidth]{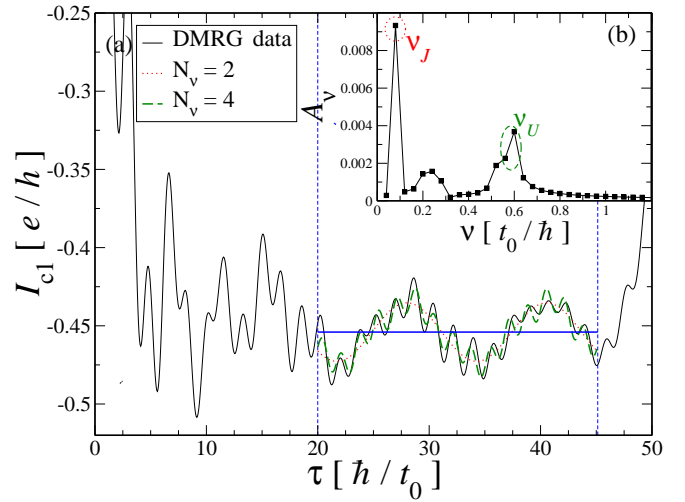}    
    \caption{(Color online) 
    (a) Black continuous line: left-contact current $I_{c1}$ for a system of $L=100$ with two impurities at distance $R=7$, $t_{\rm C}=0.8 t_{0}$, $\Delta V=0.5 t_{0}$, and $U_{\rm C}=5 t_{0}$. Horizontal continuous straight line: zero frequency component of the DFT in the interval 
    $[20,45]$ (delimited by vertical dashed lines).
   (b): DFT of the black curve in (a). The red dotted curve in  (a) corresponds to $N_{\nu}=2$ frequencies: the zero frequency component and the Josephson frequency $\nu_{\rm J}$. The green dashed curve in (a) is found using also the frequencies framed by the dashed line in (b).}
    \label{fig:FourRec}
  \end{center}
\end{figure}
In the presence of interaction, both for one and two impurities, additional frequencies emerge. In Fig.\ \ref{fig:FourRec} we show the current on the left contact for $U_C = 5 t_0$ as an example, where additional oscillations superimposed to the Josephson oscillations (they have in this case a period $T_{\rm J} \sim 12.6\hbar/t_{0}$) are clearly visible. In order to deal with the appearence of several frequencies, we perform a discrete Fourier transform (DFT) by first identifying an interval of time where the evolution is quasi-stationary, with a duration that is an integer number of Josephson periods $T_{\rm J}$. Then we do a reconstruction of the current by picking up only the few most important frequencies from the DFT, which always include the zero frequency component (the approximant to the steady-state current), the Josephson frequency $\nu_{J}$, and the one due to interaction with the highest Fourier weight $\nu_{U}$,
as displayed in Fig.~\ref{fig:FourRec}, where the quality of such a reconstruction can be seen for two different numbers of frequencies considered.
We associate to the approximant to the steady-state current the uncertainty:
\beq
\Delta I\equiv\frac{1}{M}\sqrt{\sum_{i=1,M}\left(I(\tau_{i})-\tilde I (\tau_{i})\right)^{2}}\;,
\eeq
where $\tau_{i}$, with $i=1,M$, are the equally spaced times lying in the interval where the DFT is performed, $I(\tau_{i})$ is the current 
measured at $\tau_{i}$ and $\tilde I$ is the reconstructed current.  The uncertainty $\Delta I$ is typically within the size of the symbols in our plots. 
\par
\begin{figure}[!htb]
  \begin{center}
    \includegraphics[width=0.8\columnwidth]{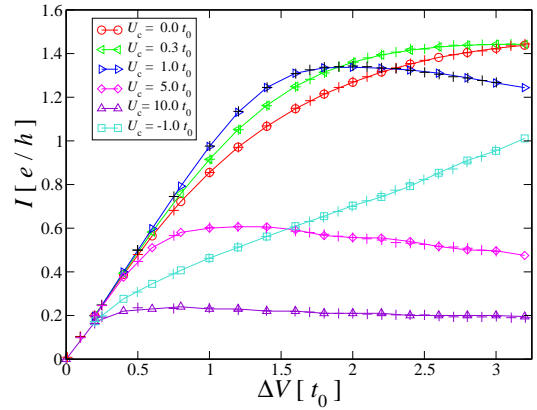}    
    \caption{(Color online) I-V characteristics of a system at $t_{\rm C}=0.5 t_{0}$ with quench scheme (B). The crosses are data from Ref.~\onlinecite{Boulat_PRL08}, the symbols are those obtained with our code (the parameters of our simulations are $L=96$, $m=600$ states, discarded weight $\varepsilon < 10^{-7}$).}
    \label{fig:compare}
  \end{center}
\end{figure}

By using the procedure described above we reproduce in Fig.~\ref{fig:compare} the I-V characteristics of a single impurity in the full range of interactions and biases with excellent agreement with the original work \cite{Boulat_PRL08}. 
%
 
\section{Results}\label{sec:results}
\subsection{Phase relations}\label{ssec:phaserelations}
As is well known, the RKKY interaction is an indirect exchange interaction between two localized spins mediated by the surrounding electrons of the Fermi sea~\cite{Ruderman_PR54,*Kasuya_PTP56,*Yosida_PR57,*VanVleck_RMP62,*Kittel}. In the present case, since we are dealing with spinless fermions, only a coupling to the density will result.
The RKKY interaction depends on the distance between the impurities $R$ via $2k_{\rm F}$ oscillations~\cite{Ruderman_PR54,*Kasuya_PTP56,*Yosida_PR57,*VanVleck_RMP62,*Kittel}
and is expected to induce correlations between the densities on the two dots and, consequently, on the currents in the contacts. We now show that the occupations on the dots 
 closely fulfill the predictions of the RKKY interaction, first considering half-filling, and then a case away from it.
The same correlations are also visible in the currents, but with a phase shift.
\par
\begin{figure}[!htb]
  \begin{center}
    \includegraphics[width=1.0\columnwidth]{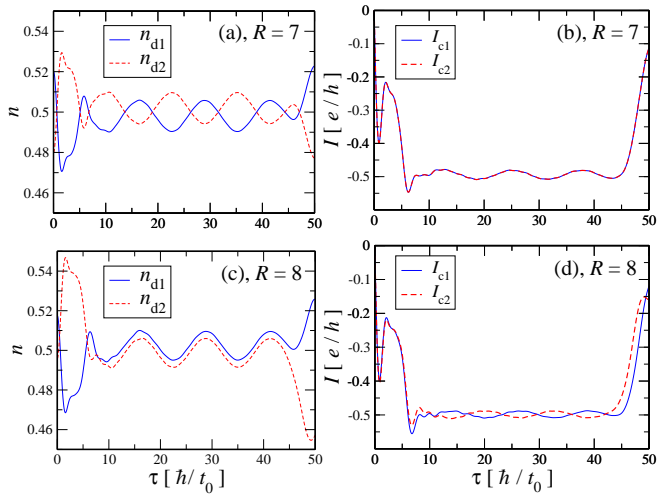}    
    \caption{(Color online) Panels (a) and (c): time evolution of the number of particles on the left (blue continuous line) and right dot (red dashed line); panels (b) and (d): time evolution of the left-contact (blue continuous line) and right-contact  (red dashed line)  currents. Data for a system of $L=100$, $t_{\rm C}=0.8 t_{0}$, $\Delta V=0.5 t_{0}$, half-filling, quench scheme (B) and $U_{\rm C}=0$.} 
    \label{fig:NvsIfree}
  \end{center}
\end{figure}
\begin{figure}[!hb]
  \begin{center}
    \includegraphics[width=0.8\columnwidth]{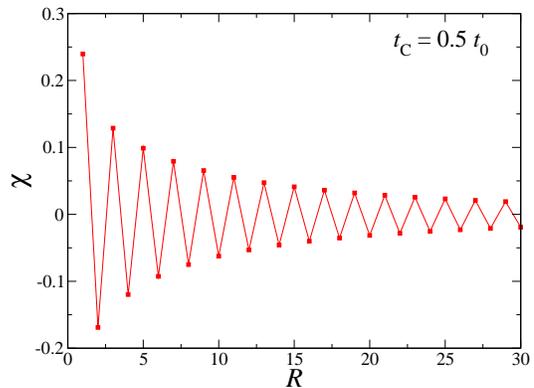}    
    \caption{(Color online) Static susceptibility connecting the dots at sites $d_1$ and $d_2$ for $t_C = 0.5 t_0$ for half-filling.} 
    \label{fig:SusceptibilityTc05}
  \end{center}
\end{figure}

We consider first 
the system at half-filling, i.e. $N/L=0.5$ in the free case and concentrate on the quasi-steady regime. In the left panels of Fig.~\ref{fig:NvsIfree} we show the occupations on the two quantum dots. 
They oscillate with the Josephson frequency
$\nu_{\rm J}$, which characterizes also the current (see Sec.~\ref{ssec:timeav}). More interestingly we observe that 
when $R$ is odd the densities oscillate in opposition of phase, while if $R$ is even they oscillate in phase. This is a regular pattern which 
we find in all the range of $R$ considered.
This behavior is compatible with the $2k_{\rm F}$ oscillations of the RKKY interaction, as shown by Fig.\ \ref{fig:SusceptibilityTc05}. 
There it can be seen that the static susceptibility, that displays $2 k_F$ oscillations as a function of $R$, is positive for $R$ odd and negative for $R$ even. Therefore, for $R$ odd the densities at the dots experience an effective repulsive interaction, while for $R$ even it is attractive.     

If we now move to the right panels of Fig.~\ref{fig:NvsIfree} we find the opposite situation. When $R$ is odd the currents
oscillate in phase (they are exactly equal in this case) and when $R$ is even they are in opposition of phase. In the latter case averaging
the currents of the two contacts cancels out the oscillations. 
This effect is visible only in the quasi-stationary regime, as we can see from the left panels of Fig.~\ref{fig:NvsIfree}.
The phase-shift between densities and currents can be undertstood 
by noticing that when the mean density on a dot increases, transfer of a particle to (from) the dot is suppressed (enhanced).  
Then, for $R$ odd, while one dot has a higher density, the other has a lower one. Considering the current on the links to the left of $d_1$ and to the right of $d_2$, charge flow is enhanced on both links when $d1$ has an increased density and $d_2$ a reduced one, while in the opposite case current is suppressed. On the other hand, when $R$ is even, both dots have an enhanced density or a suppressed one, such that when charge can be transferred on one link, the current is suppressed on the other. 
 \begin{figure}[!htb]
  \begin{center}
    \includegraphics[width=1.0\columnwidth]{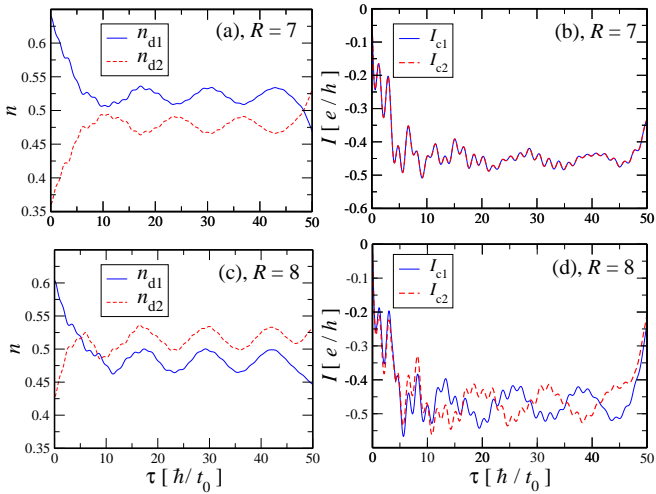}    
    \caption{(Color online) Same as Fig.~\ref{fig:NvsIfree} but for $U_{\rm C}=5 t_{0}$.} 
    \label{fig:NvsIint}
  \end{center}
\end{figure}
\par
Although the evolution of the current is 
more involved in the presence of interaction due to the appearence of additional oscillations, 
the same qualitative considerations hold also at half-filling for $U_{\rm C}\neq0$ . As an example, in Fig.~\ref{fig:NvsIint} we show the currents and the densities in the presence of interaction, namely at $U_{\rm C}=5 t_{0}$. The behavior of the densities is very clear and analogous to the free case. 
However, the 
interaction enhances the amplitude of the oscillations
as can be
seen comparing  Figs.\ \ref{fig:NvsIfree} and \ref{fig:NvsIint}. 
In spite of the interaction, 
it is clearly visible that for the odd-$R$ case the currents are exactly equal and for even $R$ 
an opposition in phase is evident. 
\par
\begin{figure}[!htb]
  \begin{center}
    \includegraphics[width=1.0\columnwidth]{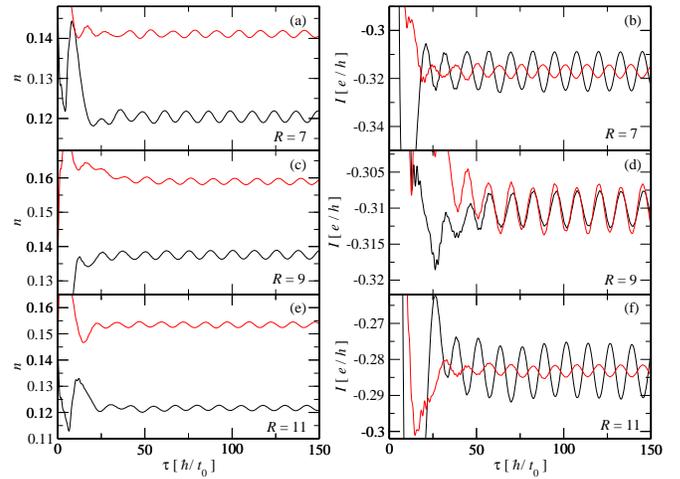}    
    \caption{(Color online) Panels (a), (c), and (e): time evolution of the number of particles on the left (black 
    line) and right dot (red line); panels (b), (d), and (f): time evolution of the left-contact (black
line) and right-contact  (red 
line)  currents. 
Data for a system of $L=400$, $t_{\rm C}=0.8 t_{0}$, $N= 98$,
 $\Delta V=0.5 t_{0}$, quench scheme (B) and $U_{\rm C}=0$.} 
    \label{fig:quafree}
  \end{center}
\end{figure}

 
Next
we consider a situation away from half filling. 
In this case, however,
already in the absence of interactions and for values of $t_{\rm C}$ different from $t_0$, the density in the central region (composed of the sites $d_{1}+1$ to $d_{2}-1$) $\rho_{c}$ does not in general coincide with $\rho=N/L$, in contrast to the
half filling case. Yet, as we discuss below, an examination of the phase differences between the densities at the quantum dots and currents across them displays a pattern that can be consistently assigned to the RKKY interaction. 
As an example we show in Fig.~\ref{fig:quafree} the density and the current for a system of $L=400$ accomodating a 
number of particles $N$ such that $\rho_{c}$, the density in the internal region, is as close as possible to quarter filling for each 
$R$  considered there. 
In particular we chose
$N=105$, which gives $\rho_{c}=0.246$ for $R=9$. 
Figures \ref{fig:quafree} (a), (c), and (e) display the oscillations of the density on each dot as a function of time. The density between them being approximately 1/4, a phase difference $\Delta \phi \simeq 3 \pi/2$  is expected, while the actual value is 1.23 $\pi$. Such a deviation corresponds to a departure of the mean density in that region of around 10\%. In spite of the slight deviation from the expected value of the phase difference for a given $R$, the periodicity four expected from the RKKY susceptibility at $k_{\rm F}=\pi/4$ is indeed found on going from $R=7$ to $R=11$ ($\Delta \phi \simeq 1.26 \pi$). This fact is, moreover, clearly seen on Figs.\ \ref{fig:quafree} (b), (d), and (f), where the current through the dots is plotted.
\par
\begin{figure}[!htb]
  \begin{center}
   \includegraphics[width=1.0\columnwidth]{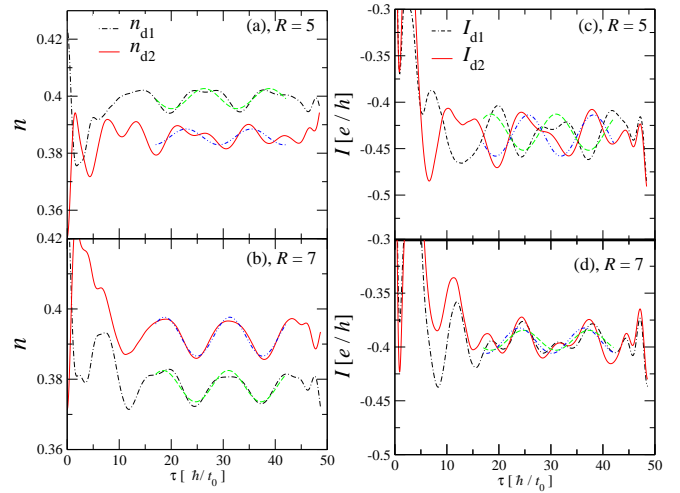}    
   \caption{(Color online) Panels (a) and (c): time evolution of the number of particles on the left (black dash-dotted)
    line) and right dot (red  
    line); panels (b) and (d): time evolution of the left-contact (black dash-dotted
line) and right-contact  (red 
line)  currents. The blue dash-dash-dotted and the green dashed lines in each panel are the reconstructed quanties, obtained retaining only
the component of the DFT in the interval $\tau\in[17,42.13]\hbar/t_{0} $ corresponding to Josephson frequency. 
Data for a system of $L=100$, $t_{\rm C}=0.8 t_{0}$, $N= 24$, $\Delta V=0.5 t_{0}$, quench scheme (B) and $U_{\rm C}=2.0t_{0}$.
}
    \label{fig:intquatime}
  \end{center}
\end{figure}

In the interacting case the presence of additional frequencies has to be taken into account, as already discussed for half-filling. Moreover, we have to consider also 
the departure of $\rho_{c}$ from $\rho$. In Fig.~\ref{fig:intquatime} we show an example of the instantaneous densities and currents with $U_{\rm C}=1.0 t_{0}$. In order to tune $\rho_{c}$ as close as possible to quarter
filling, we chose to work with $N=24$ particles, giving $\rho_{c}=0.252$ and $\rho_{c}=0.248$ for $R=5$ and $R=7$ respectively.
Performing a discrete Fourier transform on an integer number of Josephson periods (also considering different choices of the initial and final times), 
we computed the reconstructed densities and currents using only the Josephson frequency. For the time interval shown in 
Fig.~\ref{fig:intquatime} the phase between the densities and the currents changes by roughly $\pi$ going from $R$ to $R+2$, in reasonable agreement with the free case. 
However, due to the difficulty to fix the density in the central region, the results away from half-filling do not allow for a clear identification of 
phase changes as expected on the basis of the RKKY interaction. 
\par

\begin{figure}[!htb]
  \begin{center}
    \includegraphics[width=1.0\columnwidth]{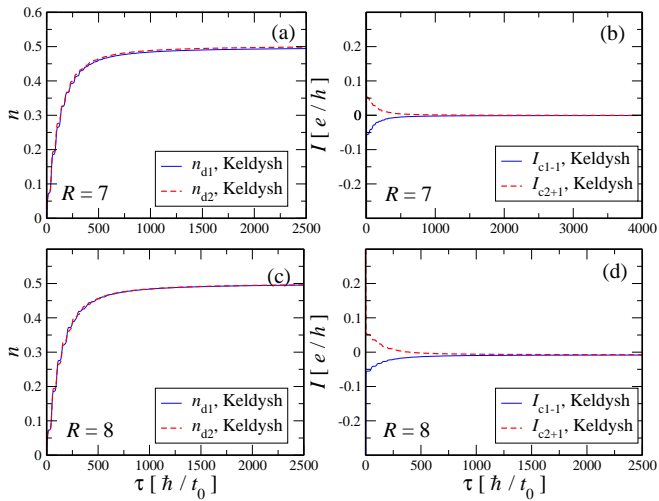}    
    \caption{(Color online) Panels (a) and (c): time evolution of the number of particles computed within the Keldysh formalism on the left (blue continuous line) and right dot (red dashed line); panels (b) and (d): time evolution of the left-lead (blue continuous line) and right-lead  (red dashed line)  currents. Data for $t_{\rm C}=0.1 t_{0}$, $t'_{\rm C}=0.12 t_{0}$ (see main text) and $\Delta V= 0.01 t_{0}$. Panels (a) and (b): $R=7$, panels (c) and (d): $R=8$.}
    \label{fig:keldysh}
  \end{center}
\end{figure}

The previous results were obtained on systems where the leads are finite, and hence allowed for a change in density. It would be on the other hand interesting to see, how much the
results change in the limit of macroscopic leads. While, as shown in Fig.\ \ref{fig:jos2}, it should be expected that the Josephson oscillations vanish, macroscopic leads will provide also an essentially infinite reservoir of fermions. It is therefore interesting to see how such reservoirs affect the region between the dots. Although it is not possible to answer this question numerically, we can obtain an insight by considering a non interacting system 
with infinite leads within the 
Keldysh formalism ~\cite{Rammer,Haug,Kennes_PRB12b} in the wide-band limit, where the density of states in the leads is considered constant. 
Following Kennes et al.~\cite{Kennes_PRB12b}, we choose a quench scheme where the bias 
is always present, the coupling to the leads is switched on at time $\tau=0$ and the sites in the central region
are initially empty. The wide-band limit is reached taking both the hopping in the contacts and the bias much smaller than the hopping in the leads, so $t_{\rm C}\ll t_{0}$ and $\Delta V\ll t_{0}$. For the Fermi sea, instead of $t_{0}$, we take a hopping $t'_{\rm C}$ of the same order of $t_{\rm C}$, in particular 
we choose $t_{\rm C}=0.1 t_{0}$ and $t'_{\rm C}=0.12 t_{0}$ and $\Delta V=0.01 t_{0}$.
In Fig.~\ref{fig:keldysh} we show our results for two impurities obtained within the Keldysh formalism, where we compute the current 
leaving the left lead ($I_{c1-1}$) and entering the right lead ($I_{c2+1}$).
We have also checked (not shown) that we obtain 
an excellent approximation of the curves shown in  Fig.~\ref{fig:keldysh}  by taking tight-binding leads in system of large size  with the 
same values of $t_{\rm C}$, $t'_{\rm C}$ and $\Delta V$, as long as the time $\tau$ is smaller than the reflection time (see Sec.~\ref{ssec:timeav}).
Both the occupations on the dots and the currents shown in Fig.~\ref{fig:keldysh} reach their steady state value after a transient and, as expected, 
no Josephson oscillations are observed. 
In both cases, $R=7$ and $R=8$,  the steady-state value of the density is close to half-filling, independently of the initial filling and of the size of the region between the dots.
Therefore, in the wide-band limit, the relevant case turns out to be that of half-filling.
The relaxation of the density to the steady state can be fitted  with an exponential of the form $n(\tau)=n_{0}(1-\exp(-\alpha \tau))$
giving  $\alpha=E_{\rm WB} / 6.1$ and $\alpha=E_{\rm WB} /8.3$ for $R=7$  and $R=8$ respectively, 
where we have introduced
the energy scale~\cite{Kennes_PRB12b} $E_{\rm WB}\equiv 4 t_{\rm C}^{2}/t_{0}$.  The time scale $1/E_{\rm WB}$ dictates the exponential relaxation of the single impurity i.e. $n(\tau)=0.5 (1- \exp(- E_{\rm WB}\tau))$,
and with  $t_{\rm C}=0.1 t_{0}$ has the value $1/E_{\rm WB}=25 \hbar/t_{0}$. If we now consider the tight-binding case of  Fig.~\ref{fig:NvsIfree}, corresponding to $t_{\rm C}=0.8$,
 we find $1/E_{\rm WB}=0.39 \hbar/t_{0}$. This is precisely the time scale over which the currents of Fig.~\ref{fig:NvsIfree} ramp from zero to the quasi-steady state regime where we 
 observe the Josephson oscillations. 
 Moving back to Fig.~\ref{fig:keldysh}, the values of the steady-state current are $I\sim 6\cdot10^{-4} e/h$ for $R=7$ and $I\sim 8 \cdot10^{-3} e/h$ for $R=8$. 
 Although these values differ significantly from each other, 
this difference is negligible with respect to the uncertainty with which the current can be accessed for example in the case of Fig.~\ref{fig:NvsIfree}. 
The smaller order of magnitude of the steady-state currents of Fig.~\ref{fig:keldysh} with respect to those of Fig.~\ref{fig:NvsIfree} 
can be understood taking into account that the bias is fifty times smaller here and also $t_{\rm C}\ll t_{0}$. 
\\
The phases characterizing the time evolution of the density and of the current described above reveal the effect of the RKKY interaction on the slow dynamics of the Josephson oscillations, giving rise to 
sustained and controllable oscillations of the densities and of the currents in finite size systems at half-filling. This fact may turn out to be observable in experiments focused on quantum dots set-ups
in mesoscopic systems. Indeed there have been proposals of  simulating quantum impurity systems and transport properties in cold atom systems~\cite{Recati_PRL05,Knap_PRX12}. The first experimental progress done so far in this direction is the realization of  a mesoscopic conducting channel in a cloud of Litihum atoms, performed by Brantut and collaborators \cite{Brantut_SCI12}. 
\subsection{Average current as a function of the distance}\label{ssec:Rdependence}
 \subsubsection{Free case}
\begin{figure}[!htb]
  \begin{center}
    \includegraphics[width=1.0\columnwidth]{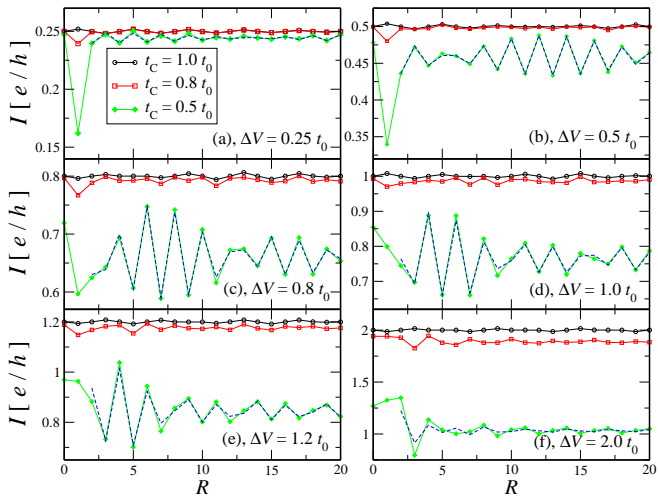}    
    \caption{(Color online) Approximant to the steady state current as a function of the distance between noninteracting impurities  for a system of $L=200$ sites for different values of the contact
    hopping, quench scheme (B) and $U_{\rm C}=0.0$. Blue dashed line: Landauer-B\"uttiker formula (cf. Eq.~\ref{eq:lb}) applied 
    to the case $t_{\rm C}=0.5 t_{0}$. From top to bottom, left to right: $\Delta V=0.25, 0.5, 0.8, 1.0, 1.2, 2.0 t_{0}$.  (The current is plotted in absolute value.).}
     \label{fig:IRfree}
  \end{center}
\end{figure}

The results shown above indicate that the dynamics of the current and the density is regulated by $2k_{\rm F}$ oscillations due to the RKKY interaction.  We now investigate how the behavior of the steady-state current
is affected by the distance between the impurities and the Fermi momentum.
In Fig.~\ref{fig:IRfree} we show the approximant to the steady-state current in absence of interaction as a function of the distance between the impurities.
The simplest case is $t_{\rm C}=t_{0}$, for which the data of Fig.~\ref{fig:IRfree} show very small variations as a function of $R$, which however are only a finite-size effect.
On the other side,  from Fig.~\ref{fig:IRfree} we see  that the curves with $t_{\rm C}=0.5 t_{0}$ are the most sensitive to $R$, showing pronounced fluctuations. 
Furthermore, 
for $t_{\rm C}=0.5 t_{0}$, there is a range of $R$ where the oscillations have the largest amplitude. This range changes with $\Delta V$. As an example, for $\Delta V=0.5 t_{0}$ the range is given approximately by $R\sim 7\div 19$, while for $\Delta V=0.8 t_{0}$ by $R\sim 4\div 13$. Moreover, the period of these oscillations is typically $R=2$. 
\par
Following the Landauer-B\"uttiker approach~\cite{Buettiker_PRL86,Blanter_PR00},
we now show that the patterns of the current of Fig.~\ref{fig:IRfree} can be understood in terms of the transmission properties
for a single particle. Indeed, the physical mechanism at the root of the flow of current is that the dot, characterized by $t_{\rm C}\neq t_{0}$,
is an effective tunnel barrier with an energy-dependent transmission probability $p_{s}(\epsilon)$ (where the subscript $s$ stands for single). The presence of two dots requires the combination of the transmission propabilities in order to compute the total probability $p_{d}(\epsilon)$ (the subscript $d$ standing for double).
The transmission probability through a single dot is given by~\cite{Branschaedel_PRB10}:
\beq\label{eq:prob}
p_{s}(\epsilon)=\frac{ 1-\epsilon^{2}/(4t_{0}^{2})}{ 1 + \epsilon^{2}(t_{0}^{2}-2t_{\rm C}^{2})/(4t_{\rm C}^{4})}\;.
\eeq
The total transmission probability can be obtained using the transfer matrix approach~\cite{Molina_EPJB05,Blanter_PR00} and gives:
\beq\label{eq:comb}
p_{d}=\frac{p_{s}^{2}}{1+(1-p_{s})^{2}-2(1-p_{s})\cos(2k(\epsilon)R+2\phi)}\;,
\eeq
where $\phi=kb$ and $b$ is the size of the single tunnel barrier. In our case we have that $t_{\rm C}$ is present on three sites (the dot
and its nearest-neighbors), so $b=3$. The expression for the combined probability eq.\ (\ref{eq:comb}) 
is valid provided $R\geq3$. Indeed, for $R=1,2$ one has to consider a single barrier of size $b=4,5$ respectively.  
In order to obtain
the average current, one has to integrate the transmission probability over the energies of
current-carrying states.  This yields~\cite{Buettiker_PRL86}:
\beq\label{eq:lb}
I(\Delta V)=\int_{\epsilon_{\rm F}-\Delta V/2}^{\epsilon_{\rm F}+\Delta V/2}d\epsilon\; p_{d}(\epsilon)\;.
\eeq
Our results for $t_{\rm C}=0.5 t_{0}$ are the blue dashed lines of Fig.~\ref{fig:IRfree}. We observe that there is a very good agreement 
with the current obtained by doing the time average. 
\par
\begin{figure}[!htb]
  \begin{center}
    \includegraphics[width=1.0\columnwidth]{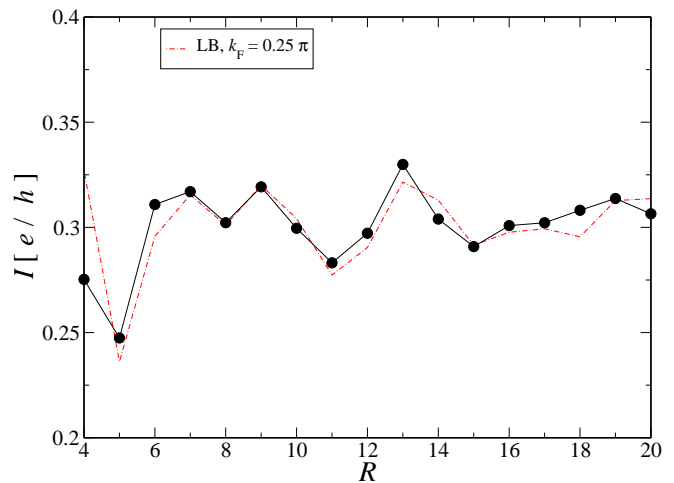}    
    \caption{(Color online) 
Approximant to the steady state current for a system with $t_{\rm C}=0.8 t_{0}$, $\Delta V=0.5 t_{0}$ and scheme (B) 
for $L=400$ (black dots) and Landauer-B\"uttiker prediction from Eq.~\ref{eq:lb} for $k_{\rm F}=0.25 \pi
$. 
(red dashed line). }
     \label{fig:0pfour}
  \end{center}
\end{figure}

In 
Fig.~\ref{fig:0pfour} we consider a system
with filling $\rho_{c}\simeq 0.25$ and show the approximant to the steady-state current (computed for a system of $L=400$ sites) and the prediction of Eq.~\ref{eq:lb}
for $k_{\rm F}=0.25.$
In spite of the difficulties in setting a definite density in the region between the dots away from half filling, a rather good agreement with the Landauer-B\"uttiker formula is obtained, 
with small deviations due to fluctuations of the density in the central region on going from one value of $R$ to another. 

\subsubsection{Interacting case}
\begin{figure}[!htb]
  \begin{center}
    \includegraphics[width=0.8\columnwidth]{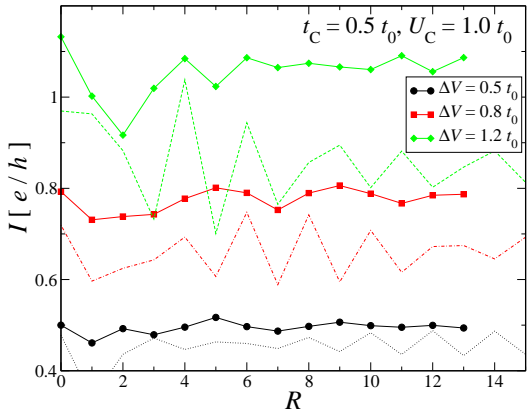}    
    \caption{(Color online) Approximant to the steady-state current as a function of the distance between the impurities $R$ for a system of $L=100$ sites with $U_{\rm C}=1.0 t_{0}$ for different values of the bias and quench scheme (B). The currents in the non-interacting case (dotted, dash-dotted and dashed lines show the current  and correspond to $\Delta V=0.5, 0.8,1.2 t_{0}$ respectively)  are shown as reference. The uncertainty on the value of the average current  $\Delta I$ as discussed in Sec.~\ref{ssec:timeav} is within the size of the symbols in the plot.  (The current is plotted in absolute value.)}
     \label{fig:IRsmallUc}
  \end{center}
\end{figure}

We start by considering the effect of a small interaction, namely $U_{\rm C}=1.0 t_{0}$, and we choose $t_{\rm C}=0.5 t_{0}$ in order to probe if and how the interaction affects the resonances 
(Fig.~\ref{fig:IRsmallUc}).
From the comparison with the free case we can see
first 
that 
the current is enhanced, an effect that becomes stronger at larger values of the bias.
The enhancement of the current by interaction is also observed
in the one impurity case~\cite{Boulat_PRL08} (see 
Fig.~\ref{fig:compare} for small $U_{\rm C}\lesssim t_{0}$ and $\Delta V\lesssim 2 t_{0}$). 
Furthermore, it can be seen
that the resonances observed in the free case
are suppressed.  The deviation of the conductance from the Landauer-B\"uttiker combination of probabilities for small values 
of the interaction was already observed in  Ref.\onlinecite{Molina_EPJB05}.
\par
\begin{figure}[!htb]
  \begin{center}
    \includegraphics[width=0.8\columnwidth]{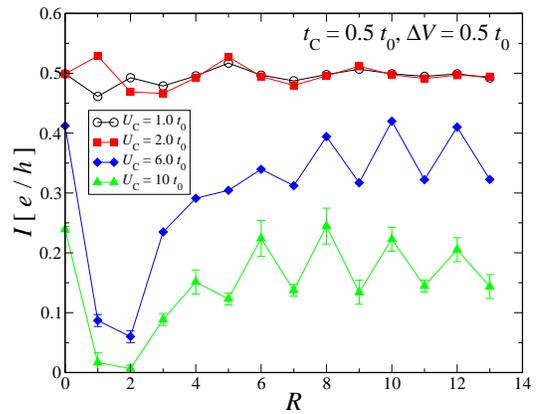}    
    \caption{(Color online) Approximant to the steady-state current as a function of the distance between the impurities $R$ for a system of $L=100$ sites with $t_{\rm C}=0.5 t_{0}$ and $\Delta V=0.5 t_{0}$ for different values of the interaction and quench scheme (B).  
    In some cases the approximant to the steady-state current is strongly sensitive on the chosen boundaries of the stationary regime
    where the DFT is performed. 
    The resulting different values of  the current lie inside the error bars.
    (The current is plotted in absolute value.)}
     \label{fig:IRsmallDV0p5}
  \end{center}
\end{figure}
\begin{figure}[!hb]
  \begin{center}
    \includegraphics[width=0.8\columnwidth]{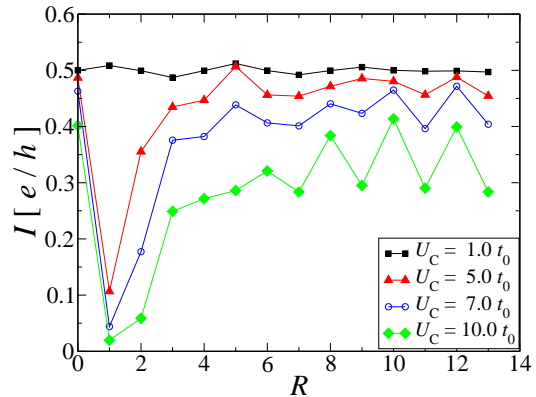}    
    \caption{(Color online) Approximant to the steady-state current as a function of the distance between the impurities $R$, in the presence of a small bias $\Delta V=0.5 t_{0}$ for a system of $L=100$ sites for different values of the interaction and $t_{\rm C}=0.8 t_{0}$. 
    The uncertainty on the value of the average current  $\Delta I$ as discussed in Sec.~\ref{ssec:timeav} is within the size of the symbols in the plot. 
       (The current is plotted in absolute value.)} 
    \label{fig:IRsmallDV}
  \end{center}
\end{figure}

In Fig.~\ref{fig:IRsmallDV0p5} we show our results for the approximant to the steady-state current with increasing values of the interaction. 
While the current does not vary significantly 
for
values of $U_C$ lower than the band-width,
a qualitatively different behavior appears when $U_C$ is larger than $ 4 t_0$. 
For $U_{\rm C}=6, 10 t_{0}$ and $R\gtrsim 6$ we interestingly find that the current oscillates as a function of 
the distance with periodicity two, with a rather large amplitude, which is typical of RKKY oscillations at half-filling.
The same behavior is also confirmed if we change the contact hopping, for example with $t_{\rm C}=0.8 t_{0}$  (Fig.\ \ref{fig:IRsmallDV}).
In Fig.~\ref{fig:IRfree} we saw that without interaction  the current is in this case almost independent on $R$, because the single tunnel barrier 
has a transmission coefficient close to unity (see Eq.~\ref{eq:prob}). 
On the contrary,
comparing Figs.~\ref{fig:IRsmallDV0p5} and ~\ref{fig:IRsmallDV} we see
that the approximant to the steady-state current oscillates with $R$ for both values of $t_C$ with
the same pattern if the interaction is large enough, i.e. $U_{\rm C}\gtrsim 5.0 t_{0}$, hinting at 
a signature of the RKKY interaction. It is also remarkable that the maxima of the current are of the same order as for the single-impurity case.
It is to be emphasized however, that even-odd oscillations of the conductance
have also been observed in a system with an impurity separated  by a 
non-interacting lead~\cite{Weinmann_EPJB08} from a non-interacting potential scatterer.

\par 
To test the dependence of
the oscillations as a function of $R$ 
on filling would be in principle desirable.
However, at low filling 
the current is drastically suppressed in the presence of large interactions. Indeed, by considering for example quarter filling with $L=100$ sites, already at $U_{\rm C}=5.0t_{0}$
the current is characterized by high-frequency oscillations around zero (data not shown), thus 
precluding  the observation of possible RKKY oscillations. 
Recalling also the problem of the deviation of $\rho$ from $\rho_{c}$ discussed in the previous sections, 
regimes away from half-filling would require the investigation of  much larger sizes, in order both to control precisely $\rho_{c}$ and to avoid strong finite-size effects present in very dilute systems with large $U_{\rm C}$, beyond the present computational capabilities.
\subsection{I-V characteristics}\label{ssec:Negative}
\begin{figure}[!hb]
  \begin{center}
    \includegraphics[width=0.8\columnwidth]{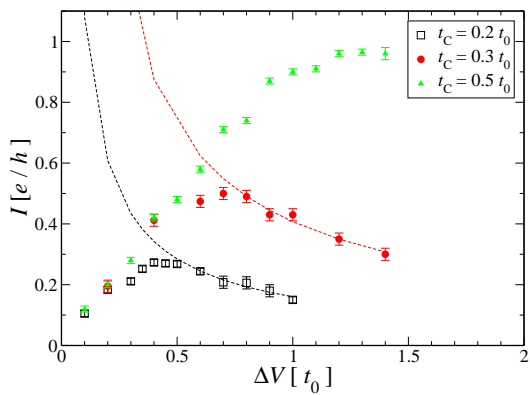}    
    \caption{(Color online) Approximant to the steady-state current as a function of the bias $\Delta V$, with $U_{\rm C}=2.0$, $R=7$ and $L=100$ sites for different values of the contact hopping $t_{\rm C}$.  Black empty squares, red full dots, green triangles correspond to $t_{\rm C}=0.2, 0.3,0.5 t_{0}$ respectively, while dashed lines are power-law fits. 
        (The current is plotted in absolute value.)} 
    \label{fig:negcond}
  \end{center}
\end{figure}

For the case of one impurity, the I-V characteristics is characterized by a regime of negative conductance, where the current decreases as 
a power-law, with interaction dependent exponents~\cite{Boulat_PRL08} (see also Fig.~\ref{fig:compare}). Furthermore, it is possible to define
an universal energy scale $T_{\rm B}$~\cite{Boulat_PRL08}, which depends on $t_{\rm C}$. At the self-dual point it gives rise to a universal power-law decay, i.e., by rescaling different
I-V characterstics with $T_{\rm B}$ they all sit on the same curve~\cite{Boulat_PRL08}. 
In the case of two impurities we also find a regime of negative conductance, as we show in Fig.~\ref{fig:negcond}. 
We observe that the behavior of the current is in some cases not very smooth. This is due to finite-size effects. 
The curves of Fig.~\ref{fig:negcond} show that the current first increases approximately linearly, has a maximum and then decreases. 
However, in order to verify if a power law may describe the sector with a negative conductance, as in the case of a single quantum dot \cite{Boulat_PRB08}, an extended range in values of the bias are necessary. In the case of two dots coupled by a Fermi sea, such a range in values of $\Delta V$ becomes very demanding in terms of the number of DMRG states that have to be kept for a reasonalbe accuracy, such that a quantitative answer cannot be given to this question.   
\section{Summary}\label{sec:conclusions}
By 
studying the time dependence of
the current and the density in a one-dimensional chain in the presence of two interacting resonant levels, we tested the interplay of the RKKY interaction and the characteristics of the quantum dots, concerning the dynamical behavior in a finite system as well as the approximant of the steady-state current.
\par
Focusing on the time evolution, 
we 
found that, at finite size, the evolution of the  current in the contacts and the occupations of the dots are characterized by oscillations, whose 
period
depends on the applied bias as in the single dot case \cite{Branschadel_AP10},
but interrelated in a way that depends
on the size of the Fermi sea. 
In fact, we
show that  the densities
on the dots oscillate 
with a relative phase which depends on the Fermi momentum of the Fermi sea and on the distance between the impurities,
as expected for
the RKKY interaction.
An analogous behavior is found for the time evolution of the currents in the contacts, which are  related to those of the 
density, but 
phase shifted with respect to them. 
While at half-filling those correlations can be clearly seen,  
away from half-filling it is necessary to precisely control the $k_{F}$  by appropriately  tuning the global density $\rho$, since the latter does not coincide in general with 
the density in the central region $\rho_{c}$, rendering the comparison for different values of $R$ difficult.
The phase relations described above can
be exploited in experimental measurements 
in mesoscopic systems.
As mentioned before, 
experimental investigation of transport in cold atomic systems~\cite{Brantut_SCI12,Killi_PRA12,Killi2_PRA12,Knap_PRX12},
would be an interesting set-up, where the variations of the density in the quantum dots could be accessed directly. 
In the thermodynamic limit the oscillations of the current and the density vanish, as we have shown 
by an explicit extrapolation, and with analytic calculations
in the wide-band limit.
\par
We have also studied the approximant to the steady state current, 
and its
oscillations
as a function of the distance between the dots.
In the free case  we identified
resonances that
can be traced back to the resonances affecting
the transmission coefficients of a single particle propagating freely in the system. 
Turning interactions on the resonances are suppressed.
However,
for large values of the interaction we observe at half-filling rather large oscillations of the current as a function of the distance with periodicity two. 
This matches  $2k_{\rm F}$ oscillations,
hinting at the influence of the
RKKY interaction.
Finally, we focused on the I-V characteristics , finding a region of
negative conductance, in analogy with 
the one-impurity case.
\acknowledgments
We thank  T. Caneva,  J. Carmelo, S. Costamagna, D. Kennes, J. Kroha,  S. Montangero and D. Rossini for useful discussions.
We acknowledge financial support from DPG through project SFB/TRR21 and Juropa/J\"ulich for the 
generous allocation of computational time.

\bibliographystyle{apsrev4-1}
\bibliography{refqd}

\begin{thebibliography}{49}%
\makeatletter
\providecommand \@ifxundefined [1]{%
 \@ifx{#1\undefined}
}%
\providecommand \@ifnum [1]{%
 \ifnum #1\expandafter \@firstoftwo
 \else \expandafter \@secondoftwo
 \fi
}%
\providecommand \@ifx [1]{%
 \ifx #1\expandafter \@firstoftwo
 \else \expandafter \@secondoftwo
 \fi
}%
\providecommand \natexlab [1]{#1}%
\providecommand \enquote  [1]{``#1''}%
\providecommand \bibnamefont  [1]{#1}%
\providecommand \bibfnamefont [1]{#1}%
\providecommand \citenamefont [1]{#1}%
\providecommand \href@noop [0]{\@secondoftwo}%
\providecommand \href [0]{\begingroup \@sanitize@url \@href}%
\providecommand \@href[1]{\@@startlink{#1}\@@href}%
\providecommand \@@href[1]{\endgroup#1\@@endlink}%
\providecommand \@sanitize@url [0]{\catcode `\\12\catcode `\$12\catcode
  `\&12\catcode `\#12\catcode `\^12\catcode `\_12\catcode `\%12\relax}%
\providecommand \@@startlink[1]{}%
\providecommand \@@endlink[0]{}%
\providecommand \url  [0]{\begingroup\@sanitize@url \@url }%
\providecommand \@url [1]{\endgroup\@href {#1}{\urlprefix }}%
\providecommand \urlprefix  [0]{URL }%
\providecommand \Eprint [0]{\href }%
\providecommand \doibase [0]{http://dx.doi.org/}%
\providecommand \selectlanguage [0]{\@gobble}%
\providecommand \bibinfo  [0]{\@secondoftwo}%
\providecommand \bibfield  [0]{\@secondoftwo}%
\providecommand \translation [1]{[#1]}%
\providecommand \BibitemOpen [0]{}%
\providecommand \bibitemStop [0]{}%
\providecommand \bibitemNoStop [0]{.\EOS\space}%
\providecommand \EOS [0]{\spacefactor3000\relax}%
\providecommand \BibitemShut  [1]{\csname bibitem#1\endcsname}%
\let\auto@bib@innerbib\@empty
\bibitem [{\citenamefont {Hewson}(1993)}]{Hewson}%
  \BibitemOpen
  \bibfield  {author} {\bibinfo {author} {\bibfnamefont {A.~C.}\ \bibnamefont
  {Hewson}},\ }\href@noop {} {\emph {\bibinfo {title} {The Kondo Problem to
  Heavy Fermions}}}\ (\bibinfo  {publisher} {Cambridge University Press},\
  \bibinfo {address} {Cambridge, UK},\ \bibinfo {year} {1993})\BibitemShut
  {NoStop}%
\bibitem [{\citenamefont {Goldhaber-Gordon}\ \emph {et~al.}(1998)\citenamefont
  {Goldhaber-Gordon}, \citenamefont {Shtrikman}, \citenamefont {Mahalu},
  \citenamefont {Abusch-Magder}, \citenamefont {Meirav},\ and\ \citenamefont
  {Kastner}}]{GoldhaberGordon_NAT98}%
  \BibitemOpen
  \bibfield  {author} {\bibinfo {author} {\bibfnamefont {D.}~\bibnamefont
  {Goldhaber-Gordon}}, \bibinfo {author} {\bibfnamefont {H.}~\bibnamefont
  {Shtrikman}}, \bibinfo {author} {\bibfnamefont {D.}~\bibnamefont {Mahalu}},
  \bibinfo {author} {\bibfnamefont {D.}~\bibnamefont {Abusch-Magder}}, \bibinfo
  {author} {\bibfnamefont {U.}~\bibnamefont {Meirav}}, \ and\ \bibinfo {author}
  {\bibfnamefont {M.}~\bibnamefont {Kastner}},\ }\href@noop {} {\bibfield
  {journal} {\bibinfo  {journal} {Nature}\ }\textbf {\bibinfo {volume} {391}},\
  \bibinfo {pages} {156} (\bibinfo {year} {1998})}\BibitemShut {NoStop}%
\bibitem [{\citenamefont {van~der Wiel}\ \emph {et~al.}(2000)\citenamefont
  {van~der Wiel}, \citenamefont {Franceschi}, \citenamefont {Fujisawa},
  \citenamefont {Elzerman}, \citenamefont {Tarucha},\ and\ \citenamefont
  {Kouwenhoven}}]{vanderWiel_SCI00}%
  \BibitemOpen
  \bibfield  {author} {\bibinfo {author} {\bibfnamefont {W.~G.}\ \bibnamefont
  {van~der Wiel}}, \bibinfo {author} {\bibfnamefont {S.~D.}\ \bibnamefont
  {Franceschi}}, \bibinfo {author} {\bibfnamefont {T.}~\bibnamefont
  {Fujisawa}}, \bibinfo {author} {\bibfnamefont {J.~M.}\ \bibnamefont
  {Elzerman}}, \bibinfo {author} {\bibfnamefont {S.}~\bibnamefont {Tarucha}}, \
  and\ \bibinfo {author} {\bibfnamefont {L.~P.}\ \bibnamefont {Kouwenhoven}},\
  }\href {\doibase 10.1126/science.289.5487.2105} {\bibfield  {journal}
  {\bibinfo  {journal} {Science}\ }\textbf {\bibinfo {volume} {289}},\ \bibinfo
  {pages} {2105} (\bibinfo {year} {2000})}\BibitemShut {NoStop}%
\bibitem [{\citenamefont {Ruderman}\ and\ \citenamefont
  {Kittel}(1954)}]{Ruderman_PR54}%
  \BibitemOpen
  \bibfield  {author} {\bibinfo {author} {\bibfnamefont {M.~A.}\ \bibnamefont
  {Ruderman}}\ and\ \bibinfo {author} {\bibfnamefont {C.}~\bibnamefont
  {Kittel}},\ }\href {\doibase 10.1103/PhysRev.96.99} {\bibfield  {journal}
  {\bibinfo  {journal} {Phys. Rev.}\ }\textbf {\bibinfo {volume} {96}},\
  \bibinfo {pages} {99} (\bibinfo {year} {1954})}\BibitemShut {NoStop}%
\bibitem [{\citenamefont {Kasuya}(1956)}]{Kasuya_PTP56}%
  \BibitemOpen
  \bibfield  {author} {\bibinfo {author} {\bibfnamefont {T.}~\bibnamefont
  {Kasuya}},\ }\href {\doibase 10.1143/PTP.16.45} {\bibfield  {journal}
  {\bibinfo  {journal} {Progress of Theoretical Physics}\ }\textbf {\bibinfo
  {volume} {16}},\ \bibinfo {pages} {45} (\bibinfo {year} {1956})}\BibitemShut
  {NoStop}%
\bibitem [{\citenamefont {Yosida}(1957)}]{Yosida_PR57}%
  \BibitemOpen
  \bibfield  {author} {\bibinfo {author} {\bibfnamefont {K.}~\bibnamefont
  {Yosida}},\ }\href {\doibase 10.1103/PhysRev.106.893} {\bibfield  {journal}
  {\bibinfo  {journal} {Phys. Rev.}\ }\textbf {\bibinfo {volume} {106}},\
  \bibinfo {pages} {893} (\bibinfo {year} {1957})}\BibitemShut {NoStop}%
\bibitem [{\citenamefont {Van~Vleck}(1962)}]{VanVleck_RMP62}%
  \BibitemOpen
  \bibfield  {author} {\bibinfo {author} {\bibfnamefont {J.~H.}\ \bibnamefont
  {Van~Vleck}},\ }\href {\doibase 10.1103/RevModPhys.34.681} {\bibfield
  {journal} {\bibinfo  {journal} {Rev. Mod. Phys.}\ }\textbf {\bibinfo {volume}
  {34}},\ \bibinfo {pages} {681} (\bibinfo {year} {1962})}\BibitemShut
  {NoStop}%
\bibitem [{\citenamefont {Kittel}(1963)}]{Kittel}%
  \BibitemOpen
  \bibfield  {author} {\bibinfo {author} {\bibfnamefont {C.}~\bibnamefont
  {Kittel}},\ }\href@noop {} {\emph {\bibinfo {title} {Quantum theory of
  solids}}}\ (\bibinfo  {publisher} {John Wiley \& Sons},\ \bibinfo {address}
  {New York, London},\ \bibinfo {year} {1963})\BibitemShut {NoStop}%
\bibitem [{\citenamefont {Jones}\ and\ \citenamefont
  {Varma}(1987)}]{Jones_PRL87}%
  \BibitemOpen
  \bibfield  {author} {\bibinfo {author} {\bibfnamefont {B.~A.}\ \bibnamefont
  {Jones}}\ and\ \bibinfo {author} {\bibfnamefont {C.~M.}\ \bibnamefont
  {Varma}},\ }\href {\doibase 10.1103/PhysRevLett.58.843} {\bibfield  {journal}
  {\bibinfo  {journal} {Phys. Rev. Lett.}\ }\textbf {\bibinfo {volume} {58}},\
  \bibinfo {pages} {843} (\bibinfo {year} {1987})}\BibitemShut {NoStop}%
\bibitem [{\citenamefont {Jones}\ \emph {et~al.}(1988)\citenamefont {Jones},
  \citenamefont {Varma},\ and\ \citenamefont {Wilkins}}]{Jones_PRL88}%
  \BibitemOpen
  \bibfield  {author} {\bibinfo {author} {\bibfnamefont {B.~A.}\ \bibnamefont
  {Jones}}, \bibinfo {author} {\bibfnamefont {C.~M.}\ \bibnamefont {Varma}}, \
  and\ \bibinfo {author} {\bibfnamefont {J.~W.}\ \bibnamefont {Wilkins}},\
  }\href {\doibase 10.1103/PhysRevLett.61.125} {\bibfield  {journal} {\bibinfo
  {journal} {Phys. Rev. Lett.}\ }\textbf {\bibinfo {volume} {61}},\ \bibinfo
  {pages} {125} (\bibinfo {year} {1988})}\BibitemShut {NoStop}%
\bibitem [{\citenamefont {Jones}\ and\ \citenamefont
  {Varma}(1989)}]{Jones_PRB89}%
  \BibitemOpen
  \bibfield  {author} {\bibinfo {author} {\bibfnamefont {B.~A.}\ \bibnamefont
  {Jones}}\ and\ \bibinfo {author} {\bibfnamefont {C.~M.}\ \bibnamefont
  {Varma}},\ }\href {\doibase 10.1103/PhysRevB.40.324} {\bibfield  {journal}
  {\bibinfo  {journal} {Phys. Rev. B}\ }\textbf {\bibinfo {volume} {40}},\
  \bibinfo {pages} {324} (\bibinfo {year} {1989})}\BibitemShut {NoStop}%
\bibitem [{\citenamefont {Affleck}\ and\ \citenamefont
  {Ludwig}(1992)}]{affleck92}%
  \BibitemOpen
  \bibfield  {author} {\bibinfo {author} {\bibfnamefont {I.}~\bibnamefont
  {Affleck}}\ and\ \bibinfo {author} {\bibfnamefont {A.~W.}\ \bibnamefont
  {Ludwig}},\ }\href@noop {} {\bibfield  {journal} {\bibinfo  {journal} {Phys.
  Rev. Lett}\ }\textbf {\bibinfo {volume} {68}},\ \bibinfo {pages} {1046}
  (\bibinfo {year} {1992})}\BibitemShut {NoStop}%
\bibitem [{\citenamefont {Affleck}\ \emph {et~al.}(1995)\citenamefont
  {Affleck}, \citenamefont {Ludwig},\ and\ \citenamefont {Jones}}]{affleck95}%
  \BibitemOpen
  \bibfield  {author} {\bibinfo {author} {\bibfnamefont {I.}~\bibnamefont
  {Affleck}}, \bibinfo {author} {\bibfnamefont {A.~W.~W.}\ \bibnamefont
  {Ludwig}}, \ and\ \bibinfo {author} {\bibfnamefont {B.~A.}\ \bibnamefont
  {Jones}},\ }\href {\doibase 10.1103/PhysRevB.52.9528} {\bibfield  {journal}
  {\bibinfo  {journal} {Phys. Rev. B}\ }\textbf {\bibinfo {volume} {52}},\
  \bibinfo {pages} {9528} (\bibinfo {year} {1995})}\BibitemShut {NoStop}%
\bibitem [{\citenamefont {Craig}\ \emph {et~al.}(2004)\citenamefont {Craig},
  \citenamefont {Taylor}, \citenamefont {Lester}, \citenamefont {Marcus},
  \citenamefont {Hanson},\ and\ \citenamefont {Gossard}}]{Craig_Sc04}%
  \BibitemOpen
  \bibfield  {author} {\bibinfo {author} {\bibfnamefont {N.~J.}\ \bibnamefont
  {Craig}}, \bibinfo {author} {\bibfnamefont {J.~M.}\ \bibnamefont {Taylor}},
  \bibinfo {author} {\bibfnamefont {E.~A.}\ \bibnamefont {Lester}}, \bibinfo
  {author} {\bibfnamefont {C.~M.}\ \bibnamefont {Marcus}}, \bibinfo {author}
  {\bibfnamefont {M.~P.}\ \bibnamefont {Hanson}}, \ and\ \bibinfo {author}
  {\bibfnamefont {A.~C.}\ \bibnamefont {Gossard}},\ }\href {\doibase
  10.1126/science.1095452} {\bibfield  {journal} {\bibinfo  {journal}
  {Science}\ }\textbf {\bibinfo {volume} {304}},\ \bibinfo {pages} {565}
  (\bibinfo {year} {2004})}\BibitemShut {NoStop}%
\bibitem [{\citenamefont {Boulat}\ \emph {et~al.}(2008)\citenamefont {Boulat},
  \citenamefont {Saleur},\ and\ \citenamefont {Schmitteckert}}]{Boulat_PRL08}%
  \BibitemOpen
  \bibfield  {author} {\bibinfo {author} {\bibfnamefont {E.}~\bibnamefont
  {Boulat}}, \bibinfo {author} {\bibfnamefont {H.}~\bibnamefont {Saleur}}, \
  and\ \bibinfo {author} {\bibfnamefont {P.}~\bibnamefont {Schmitteckert}},\
  }\href {\doibase 10.1103/PhysRevLett.101.140601} {\bibfield  {journal}
  {\bibinfo  {journal} {Phys. Rev. Lett.}\ }\textbf {\bibinfo {volume} {101}},\
  \bibinfo {pages} {140601} (\bibinfo {year} {2008})}\BibitemShut {NoStop}%
\bibitem [{\citenamefont {Karrasch}\ \emph
  {et~al.}(2010{\natexlab{a}})\citenamefont {Karrasch}, \citenamefont
  {Pletyukhov}, \citenamefont {Borda},\ and\ \citenamefont
  {Meden}}]{Karrasch_PRB10}%
  \BibitemOpen
  \bibfield  {author} {\bibinfo {author} {\bibfnamefont {C.}~\bibnamefont
  {Karrasch}}, \bibinfo {author} {\bibfnamefont {M.}~\bibnamefont
  {Pletyukhov}}, \bibinfo {author} {\bibfnamefont {L.}~\bibnamefont {Borda}}, \
  and\ \bibinfo {author} {\bibfnamefont {V.}~\bibnamefont {Meden}},\ }\href
  {\doibase 10.1103/PhysRevB.81.125122} {\bibfield  {journal} {\bibinfo
  {journal} {Phys. Rev. B}\ }\textbf {\bibinfo {volume} {81}},\ \bibinfo
  {pages} {125122} (\bibinfo {year} {2010}{\natexlab{a}})}\BibitemShut
  {NoStop}%
\bibitem [{\citenamefont {Karrasch}\ \emph
  {et~al.}(2010{\natexlab{b}})\citenamefont {Karrasch}, \citenamefont
  {Andergassen}, \citenamefont {Pletyukhov}, \citenamefont {Schuricht},
  \citenamefont {Borda}, \citenamefont {Meden},\ and\ \citenamefont
  {Schoeller}}]{Karrasch_EPL10}%
  \BibitemOpen
  \bibfield  {author} {\bibinfo {author} {\bibfnamefont {C.}~\bibnamefont
  {Karrasch}}, \bibinfo {author} {\bibfnamefont {S.}~\bibnamefont
  {Andergassen}}, \bibinfo {author} {\bibfnamefont {M.}~\bibnamefont
  {Pletyukhov}}, \bibinfo {author} {\bibfnamefont {D.}~\bibnamefont
  {Schuricht}}, \bibinfo {author} {\bibfnamefont {L.}~\bibnamefont {Borda}},
  \bibinfo {author} {\bibfnamefont {V.}~\bibnamefont {Meden}}, \ and\ \bibinfo
  {author} {\bibfnamefont {H.}~\bibnamefont {Schoeller}},\ }\href
  {http://stacks.iop.org/0295-5075/90/i=3/a=30003} {\bibfield  {journal}
  {\bibinfo  {journal} {Europhys. Lett.}\ }\textbf {\bibinfo {volume} {90}},\
  \bibinfo {pages} {30003} (\bibinfo {year} {2010}{\natexlab{b}})}\BibitemShut
  {NoStop}%
\bibitem [{\citenamefont {Kennes}\ and\ \citenamefont
  {Meden}(2012)}]{Kennes_PRB12}%
  \BibitemOpen
  \bibfield  {author} {\bibinfo {author} {\bibfnamefont {D.~M.}\ \bibnamefont
  {Kennes}}\ and\ \bibinfo {author} {\bibfnamefont {V.}~\bibnamefont {Meden}},\
  }\href {\doibase 10.1103/PhysRevB.85.245101} {\bibfield  {journal} {\bibinfo
  {journal} {Phys. Rev. B}\ }\textbf {\bibinfo {volume} {85}},\ \bibinfo
  {pages} {245101} (\bibinfo {year} {2012})}\BibitemShut {NoStop}%
\bibitem [{\citenamefont {Kennes}\ and\ \citenamefont
  {Meden}(2013)}]{Kennes_pp12a}%
  \BibitemOpen
  \bibfield  {author} {\bibinfo {author} {\bibfnamefont {D.~M.}\ \bibnamefont
  {Kennes}}\ and\ \bibinfo {author} {\bibfnamefont {V.}~\bibnamefont {Meden}},\
  }\href {\doibase 10.1103/PhysRevB.87.075130} {\bibfield  {journal} {\bibinfo
  {journal} {Phys. Rev. B}\ }\textbf {\bibinfo {volume} {87}},\ \bibinfo
  {pages} {075130} (\bibinfo {year} {2013})}\BibitemShut {NoStop}%
\bibitem [{\citenamefont {Kennes}\ \emph {et~al.}(2012)\citenamefont {Kennes},
  \citenamefont {Jakobs}, \citenamefont {Karrasch},\ and\ \citenamefont
  {Meden}}]{Kennes_PRB12b}%
  \BibitemOpen
  \bibfield  {author} {\bibinfo {author} {\bibfnamefont {D.~M.}\ \bibnamefont
  {Kennes}}, \bibinfo {author} {\bibfnamefont {S.~G.}\ \bibnamefont {Jakobs}},
  \bibinfo {author} {\bibfnamefont {C.}~\bibnamefont {Karrasch}}, \ and\
  \bibinfo {author} {\bibfnamefont {V.}~\bibnamefont {Meden}},\ }\href
  {\doibase 10.1103/PhysRevB.85.085113} {\bibfield  {journal} {\bibinfo
  {journal} {Phys. Rev. B}\ }\textbf {\bibinfo {volume} {85}},\ \bibinfo
  {pages} {085113} (\bibinfo {year} {2012})}\BibitemShut {NoStop}%
\bibitem [{\citenamefont {Andergassen}\ \emph {et~al.}(2011)\citenamefont
  {Andergassen}, \citenamefont {Pletyukhov}, \citenamefont {Schuricht},
  \citenamefont {Schoeller},\ and\ \citenamefont {Borda}}]{Andergassen_PRB11}%
  \BibitemOpen
  \bibfield  {author} {\bibinfo {author} {\bibfnamefont {S.}~\bibnamefont
  {Andergassen}}, \bibinfo {author} {\bibfnamefont {M.}~\bibnamefont
  {Pletyukhov}}, \bibinfo {author} {\bibfnamefont {D.}~\bibnamefont
  {Schuricht}}, \bibinfo {author} {\bibfnamefont {H.}~\bibnamefont
  {Schoeller}}, \ and\ \bibinfo {author} {\bibfnamefont {L.}~\bibnamefont
  {Borda}},\ }\href {\doibase 10.1103/PhysRevB.83.205103} {\bibfield  {journal}
  {\bibinfo  {journal} {Phys. Rev. B}\ }\textbf {\bibinfo {volume} {83}},\
  \bibinfo {pages} {205103} (\bibinfo {year} {2011})}\BibitemShut {NoStop}%
\bibitem [{\citenamefont
  {Schmitteckert}(2004{\natexlab{a}})}]{Schmitteckert_PRB04}%
  \BibitemOpen
  \bibfield  {author} {\bibinfo {author} {\bibfnamefont {P.}~\bibnamefont
  {Schmitteckert}},\ }\href {\doibase 10.1103/PhysRevB.70.121302} {\bibfield
  {journal} {\bibinfo  {journal} {Phys. Rev. B}\ }\textbf {\bibinfo {volume}
  {70}},\ \bibinfo {pages} {121302} (\bibinfo {year}
  {2004}{\natexlab{a}})}\BibitemShut {NoStop}%
\bibitem [{\citenamefont {Bransch\"adel}\ \emph
  {et~al.}(2010{\natexlab{a}})\citenamefont {Bransch\"adel}, \citenamefont
  {Schneider},\ and\ \citenamefont {Schmitteckert}}]{Branschadel_AP10}%
  \BibitemOpen
  \bibfield  {author} {\bibinfo {author} {\bibfnamefont {A.}~\bibnamefont
  {Bransch\"adel}}, \bibinfo {author} {\bibfnamefont {G.}~\bibnamefont
  {Schneider}}, \ and\ \bibinfo {author} {\bibfnamefont {P.}~\bibnamefont
  {Schmitteckert}},\ }\href {\doibase 10.1002/andp.201000017} {\bibfield
  {journal} {\bibinfo  {journal} {Ann. Phys. (Berlin)}\ }\textbf {\bibinfo
  {volume} {522}},\ \bibinfo {pages} {657} (\bibinfo {year}
  {2010}{\natexlab{a}})}\BibitemShut {NoStop}%
\bibitem [{\citenamefont {Einhellinger}\ \emph {et~al.}(2012)\citenamefont
  {Einhellinger}, \citenamefont {Cojuhovschi},\ and\ \citenamefont
  {Jeckelmann}}]{Einhellinger_PRB12}%
  \BibitemOpen
  \bibfield  {author} {\bibinfo {author} {\bibfnamefont {M.}~\bibnamefont
  {Einhellinger}}, \bibinfo {author} {\bibfnamefont {A.}~\bibnamefont
  {Cojuhovschi}}, \ and\ \bibinfo {author} {\bibfnamefont {E.}~\bibnamefont
  {Jeckelmann}},\ }\href {\doibase 10.1103/PhysRevB.85.235141} {\bibfield
  {journal} {\bibinfo  {journal} {Phys. Rev. B}\ }\textbf {\bibinfo {volume}
  {85}},\ \bibinfo {pages} {235141} (\bibinfo {year} {2012})}\BibitemShut
  {NoStop}%
\bibitem [{\citenamefont {Bransch\"adel}\ \emph
  {et~al.}(2010{\natexlab{b}})\citenamefont {Bransch\"adel}, \citenamefont
  {Boulat}, \citenamefont {Saleur},\ and\ \citenamefont
  {Schmitteckert}}]{Branschaedel_PRB10}%
  \BibitemOpen
  \bibfield  {author} {\bibinfo {author} {\bibfnamefont {A.}~\bibnamefont
  {Bransch\"adel}}, \bibinfo {author} {\bibfnamefont {E.}~\bibnamefont
  {Boulat}}, \bibinfo {author} {\bibfnamefont {H.}~\bibnamefont {Saleur}}, \
  and\ \bibinfo {author} {\bibfnamefont {P.}~\bibnamefont {Schmitteckert}},\
  }\href {\doibase 10.1103/PhysRevB.82.205414} {\bibfield  {journal} {\bibinfo
  {journal} {Phys. Rev. B}\ }\textbf {\bibinfo {volume} {82}},\ \bibinfo
  {pages} {205414} (\bibinfo {year} {2010}{\natexlab{b}})}\BibitemShut
  {NoStop}%
\bibitem [{\citenamefont {Bransch\"adel}\ \emph
  {et~al.}(2010{\natexlab{c}})\citenamefont {Bransch\"adel}, \citenamefont
  {Boulat}, \citenamefont {Saleur},\ and\ \citenamefont
  {Schmitteckert}}]{Branschaedel_PRL10}%
  \BibitemOpen
  \bibfield  {author} {\bibinfo {author} {\bibfnamefont {A.}~\bibnamefont
  {Bransch\"adel}}, \bibinfo {author} {\bibfnamefont {E.}~\bibnamefont
  {Boulat}}, \bibinfo {author} {\bibfnamefont {H.}~\bibnamefont {Saleur}}, \
  and\ \bibinfo {author} {\bibfnamefont {P.}~\bibnamefont {Schmitteckert}},\
  }\href {\doibase 10.1103/PhysRevLett.105.146805} {\bibfield  {journal}
  {\bibinfo  {journal} {Phys. Rev. Lett.}\ }\textbf {\bibinfo {volume} {105}},\
  \bibinfo {pages} {146805} (\bibinfo {year} {2010}{\natexlab{c}})}\BibitemShut
  {NoStop}%
\bibitem [{\citenamefont {Carr}\ \emph {et~al.}(2011)\citenamefont {Carr},
  \citenamefont {Bagrets},\ and\ \citenamefont {Schmitteckert}}]{Carr_PRL11}%
  \BibitemOpen
  \bibfield  {author} {\bibinfo {author} {\bibfnamefont {S.~T.}\ \bibnamefont
  {Carr}}, \bibinfo {author} {\bibfnamefont {D.~A.}\ \bibnamefont {Bagrets}}, \
  and\ \bibinfo {author} {\bibfnamefont {P.}~\bibnamefont {Schmitteckert}},\
  }\href {\doibase 10.1103/PhysRevLett.107.206801} {\bibfield  {journal}
  {\bibinfo  {journal} {Phys. Rev. Lett.}\ }\textbf {\bibinfo {volume} {107}},\
  \bibinfo {pages} {206801} (\bibinfo {year} {2011})}\BibitemShut {NoStop}%
\bibitem [{\citenamefont {{Schneider}}\ and\ \citenamefont
  {{Schmitteckert}}(2006)}]{Schneider_pp06}%
  \BibitemOpen
  \bibfield  {author} {\bibinfo {author} {\bibfnamefont {G.}~\bibnamefont
  {{Schneider}}}\ and\ \bibinfo {author} {\bibfnamefont {P.}~\bibnamefont
  {{Schmitteckert}}},\ }\href@noop {} {\bibfield  {journal} {\bibinfo
  {journal} {arXiv:cond-mat/0601389}\ } (\bibinfo {year} {2006})}\BibitemShut
  {NoStop}%
\bibitem [{\citenamefont {Enss}\ \emph {et~al.}(2005)\citenamefont {Enss},
  \citenamefont {Meden}, \citenamefont {Andergassen}, \citenamefont
  {Barnab\'e-Th\'eriault}, \citenamefont {Metzner},\ and\ \citenamefont
  {Sch\"onhammer}}]{Enss_PRB05}%
  \BibitemOpen
  \bibfield  {author} {\bibinfo {author} {\bibfnamefont {T.}~\bibnamefont
  {Enss}}, \bibinfo {author} {\bibfnamefont {V.}~\bibnamefont {Meden}},
  \bibinfo {author} {\bibfnamefont {S.}~\bibnamefont {Andergassen}}, \bibinfo
  {author} {\bibfnamefont {X.}~\bibnamefont {Barnab\'e-Th\'eriault}}, \bibinfo
  {author} {\bibfnamefont {W.}~\bibnamefont {Metzner}}, \ and\ \bibinfo
  {author} {\bibfnamefont {K.}~\bibnamefont {Sch\"onhammer}},\ }\href {\doibase
  10.1103/PhysRevB.71.155401} {\bibfield  {journal} {\bibinfo  {journal} {Phys.
  Rev. B}\ }\textbf {\bibinfo {volume} {71}},\ \bibinfo {pages} {155401}
  (\bibinfo {year} {2005})}\BibitemShut {NoStop}%
\bibitem [{\citenamefont {Costamagna}\ and\ \citenamefont
  {Riera}(2008)}]{Costamagna_PRB08}%
  \BibitemOpen
  \bibfield  {author} {\bibinfo {author} {\bibfnamefont {S.}~\bibnamefont
  {Costamagna}}\ and\ \bibinfo {author} {\bibfnamefont {J.~A.}\ \bibnamefont
  {Riera}},\ }\href {\doibase 10.1103/PhysRevB.77.235103} {\bibfield  {journal}
  {\bibinfo  {journal} {Phys. Rev. B}\ }\textbf {\bibinfo {volume} {77}},\
  \bibinfo {pages} {235103} (\bibinfo {year} {2008})}\BibitemShut {NoStop}%
\bibitem [{\citenamefont {{R.A. Molina}}\ \emph {et~al.}(2005)\citenamefont
  {{R.A. Molina}}, \citenamefont {{D. Weinmann}},\ and\ \citenamefont {{J.-L.
  Pichard}}}]{Molina_EPJB05}%
  \BibitemOpen
  \bibfield  {author} {\bibinfo {author} {\bibnamefont {{R.A. Molina}}},
  \bibinfo {author} {\bibnamefont {{D. Weinmann}}}, \ and\ \bibinfo {author}
  {\bibnamefont {{J.-L. Pichard}}},\ }\href {\doibase
  10.1140/epjb/e2005-00403-1} {\bibfield  {journal} {\bibinfo  {journal} {Eur.
  Phys. J. B}\ }\textbf {\bibinfo {volume} {48}},\ \bibinfo {pages} {243}
  (\bibinfo {year} {2005})}\BibitemShut {NoStop}%
\bibitem [{\citenamefont {Weinmann}\ \emph {et~al.}(2008)\citenamefont
  {Weinmann}, \citenamefont {Jalabert}, \citenamefont {Freyn}, \citenamefont
  {Ingold},\ and\ \citenamefont {Pichard}}]{Weinmann_EPJB08}%
  \BibitemOpen
  \bibfield  {author} {\bibinfo {author} {\bibfnamefont {D.}~\bibnamefont
  {Weinmann}}, \bibinfo {author} {\bibfnamefont {R.~A.}\ \bibnamefont
  {Jalabert}}, \bibinfo {author} {\bibfnamefont {A.}~\bibnamefont {Freyn}},
  \bibinfo {author} {\bibfnamefont {G.-L.}\ \bibnamefont {Ingold}}, \ and\
  \bibinfo {author} {\bibfnamefont {J.-L.}\ \bibnamefont {Pichard}},\ }\href
  {\doibase 10.1140/epjb/e2008-00403-7} {\bibfield  {journal} {\bibinfo
  {journal} {The European Physical Journal B}\ }\textbf {\bibinfo {volume}
  {66}},\ \bibinfo {pages} {239} (\bibinfo {year} {2008})}\BibitemShut
  {NoStop}%
\bibitem [{\citenamefont {White}\ and\ \citenamefont
  {Feiguin}(2004)}]{White_PRL04}%
  \BibitemOpen
  \bibfield  {author} {\bibinfo {author} {\bibfnamefont {S.~R.}\ \bibnamefont
  {White}}\ and\ \bibinfo {author} {\bibfnamefont {A.~E.}\ \bibnamefont
  {Feiguin}},\ }\href {\doibase 10.1103/PhysRevLett.93.076401} {\bibfield
  {journal} {\bibinfo  {journal} {Phys. Rev. Lett.}\ }\textbf {\bibinfo
  {volume} {93}},\ \bibinfo {pages} {076401} (\bibinfo {year}
  {2004})}\BibitemShut {NoStop}%
\bibitem [{\citenamefont {Daley}\ \emph {et~al.}(2004)\citenamefont {Daley},
  \citenamefont {Kollath}, \citenamefont {Schollw\"ock},\ and\ \citenamefont
  {Vidal}}]{Daley_JSTAT04}%
  \BibitemOpen
  \bibfield  {author} {\bibinfo {author} {\bibfnamefont {A.~J.}\ \bibnamefont
  {Daley}}, \bibinfo {author} {\bibfnamefont {C.}~\bibnamefont {Kollath}},
  \bibinfo {author} {\bibfnamefont {U.}~\bibnamefont {Schollw\"ock}}, \ and\
  \bibinfo {author} {\bibfnamefont {G.}~\bibnamefont {Vidal}},\ }\href
  {http://stacks.iop.org/1742-5468/2004/i=04/a=P04005} {\bibfield  {journal}
  {\bibinfo  {journal} {JSTAT}\ }\textbf {\bibinfo {volume} {2004}},\ \bibinfo
  {pages} {P04005} (\bibinfo {year} {2004})}\BibitemShut {NoStop}%
\bibitem [{\citenamefont
  {Schmitteckert}(2004{\natexlab{b}})}]{schmitteckert04}%
  \BibitemOpen
  \bibfield  {author} {\bibinfo {author} {\bibfnamefont {P.}~\bibnamefont
  {Schmitteckert}},\ }\href {\doibase 10.1103/PhysRevB.70.121302} {\bibfield
  {journal} {\bibinfo  {journal} {Phys. Rev. B}\ }\textbf {\bibinfo {volume}
  {70}},\ \bibinfo {pages} {121302} (\bibinfo {year}
  {2004}{\natexlab{b}})}\BibitemShut {NoStop}%
\bibitem [{\citenamefont {Schollw\"ock}(2011)}]{Schollwoeck_AP11}%
  \BibitemOpen
  \bibfield  {author} {\bibinfo {author} {\bibfnamefont {U.}~\bibnamefont
  {Schollw\"ock}},\ }\href {\doibase 10.1016/j.aop.2010.09.012} {\bibfield
  {journal} {\bibinfo  {journal} {Annals of Physics}\ }\textbf {\bibinfo
  {volume} {326}},\ \bibinfo {pages} {96 } (\bibinfo {year}
  {2011})}\BibitemShut {NoStop}%
\bibitem [{\citenamefont {Al-Hassanieh}\ \emph {et~al.}(2006)\citenamefont
  {Al-Hassanieh}, \citenamefont {Feiguin}, \citenamefont {Riera}, \citenamefont
  {B\"usser},\ and\ \citenamefont {Dagotto}}]{AH_PRB06}%
  \BibitemOpen
  \bibfield  {author} {\bibinfo {author} {\bibfnamefont {K.~A.}\ \bibnamefont
  {Al-Hassanieh}}, \bibinfo {author} {\bibfnamefont {A.~E.}\ \bibnamefont
  {Feiguin}}, \bibinfo {author} {\bibfnamefont {J.~A.}\ \bibnamefont {Riera}},
  \bibinfo {author} {\bibfnamefont {C.~A.}\ \bibnamefont {B\"usser}}, \ and\
  \bibinfo {author} {\bibfnamefont {E.}~\bibnamefont {Dagotto}},\ }\href
  {\doibase 10.1103/PhysRevB.73.195304} {\bibfield  {journal} {\bibinfo
  {journal} {Phys. Rev. B}\ }\textbf {\bibinfo {volume} {73}},\ \bibinfo
  {pages} {195304} (\bibinfo {year} {2006})}\BibitemShut {NoStop}%
\bibitem [{\citenamefont {Cini}(1980)}]{Cini_PRB80}%
  \BibitemOpen
  \bibfield  {author} {\bibinfo {author} {\bibfnamefont {M.}~\bibnamefont
  {Cini}},\ }\href {\doibase 10.1103/PhysRevB.22.5887} {\bibfield  {journal}
  {\bibinfo  {journal} {Phys. Rev. B}\ }\textbf {\bibinfo {volume} {22}},\
  \bibinfo {pages} {5887} (\bibinfo {year} {1980})}\BibitemShut {NoStop}%
\bibitem [{\citenamefont {{Nuss}}\ \emph {et~al.}(2013)\citenamefont {{Nuss}},
  \citenamefont {{Ganahl}}, \citenamefont {{Evertz}}, \citenamefont
  {{Arrigoni}},\ and\ \citenamefont {{von der Linden}}}]{Nuss_pp13}%
  \BibitemOpen
  \bibfield  {author} {\bibinfo {author} {\bibfnamefont {M.}~\bibnamefont
  {{Nuss}}}, \bibinfo {author} {\bibfnamefont {M.}~\bibnamefont {{Ganahl}}},
  \bibinfo {author} {\bibfnamefont {H.~G.}\ \bibnamefont {{Evertz}}}, \bibinfo
  {author} {\bibfnamefont {E.}~\bibnamefont {{Arrigoni}}}, \ and\ \bibinfo
  {author} {\bibfnamefont {W.}~\bibnamefont {{von der Linden}}},\ }\href@noop
  {} {\bibfield  {journal} {\bibinfo  {journal} {arXiv:1301.3068}\ } (\bibinfo
  {year} {2013})}\BibitemShut {NoStop}%
\bibitem [{\citenamefont {Rammer}(2007)}]{Rammer}%
  \BibitemOpen
  \bibfield  {author} {\bibinfo {author} {\bibfnamefont {J.}~\bibnamefont
  {Rammer}},\ }\href@noop {} {\emph {\bibinfo {title} {Quantum Field Theory of
  Non-equilibrium States}}}\ (\bibinfo  {publisher} {Cambridge University
  Press},\ \bibinfo {address} {Cambridge, UK},\ \bibinfo {year}
  {2007})\BibitemShut {NoStop}%
\bibitem [{\citenamefont {Haug}\ and\ \citenamefont {Jauho}(2008)}]{Haug}%
  \BibitemOpen
  \bibfield  {author} {\bibinfo {author} {\bibfnamefont {H.}~\bibnamefont
  {Haug}}\ and\ \bibinfo {author} {\bibfnamefont {A.-P.}\ \bibnamefont
  {Jauho}},\ }\href@noop {} {\emph {\bibinfo {title} {Quantum Kinetics in
  Transport and Optics of Semiconductors}}}\ (\bibinfo  {publisher}
  {Springer-Verlag},\ \bibinfo {address} {Berlin},\ \bibinfo {year}
  {2008})\BibitemShut {NoStop}%
\bibitem [{\citenamefont {Recati}\ \emph {et~al.}(2005)\citenamefont {Recati},
  \citenamefont {Fedichev}, \citenamefont {Zwerger}, \citenamefont {von
  Delft},\ and\ \citenamefont {Zoller}}]{Recati_PRL05}%
  \BibitemOpen
  \bibfield  {author} {\bibinfo {author} {\bibfnamefont {A.}~\bibnamefont
  {Recati}}, \bibinfo {author} {\bibfnamefont {P.~O.}\ \bibnamefont
  {Fedichev}}, \bibinfo {author} {\bibfnamefont {W.}~\bibnamefont {Zwerger}},
  \bibinfo {author} {\bibfnamefont {J.}~\bibnamefont {von Delft}}, \ and\
  \bibinfo {author} {\bibfnamefont {P.}~\bibnamefont {Zoller}},\ }\href@noop {}
  {\bibfield  {journal} {\bibinfo  {journal} {Phys. Rev. Lett.}\ }\textbf
  {\bibinfo {volume} {94}},\ \bibinfo {pages} {040404} (\bibinfo {year}
  {2005})}\BibitemShut {NoStop}%
\bibitem [{\citenamefont {Knap}\ \emph {et~al.}(2012)\citenamefont {Knap},
  \citenamefont {Shashi}, \citenamefont {Nishida}, \citenamefont {Imambekov},
  \citenamefont {Abanin},\ and\ \citenamefont {Demler}}]{Knap_PRX12}%
  \BibitemOpen
  \bibfield  {author} {\bibinfo {author} {\bibfnamefont {M.}~\bibnamefont
  {Knap}}, \bibinfo {author} {\bibfnamefont {A.}~\bibnamefont {Shashi}},
  \bibinfo {author} {\bibfnamefont {Y.}~\bibnamefont {Nishida}}, \bibinfo
  {author} {\bibfnamefont {A.}~\bibnamefont {Imambekov}}, \bibinfo {author}
  {\bibfnamefont {D.~A.}\ \bibnamefont {Abanin}}, \ and\ \bibinfo {author}
  {\bibfnamefont {E.}~\bibnamefont {Demler}},\ }\href {\doibase
  10.1103/PhysRevX.2.041020} {\bibfield  {journal} {\bibinfo  {journal} {Phys.
  Rev. X}\ }\textbf {\bibinfo {volume} {2}},\ \bibinfo {pages} {041020}
  (\bibinfo {year} {2012})}\BibitemShut {NoStop}%
\bibitem [{\citenamefont {Brantut}\ \emph {et~al.}(2012)\citenamefont
  {Brantut}, \citenamefont {Meineke}, \citenamefont {Stadler}, \citenamefont
  {Krinner},\ and\ \citenamefont {Esslinger}}]{Brantut_SCI12}%
  \BibitemOpen
  \bibfield  {author} {\bibinfo {author} {\bibfnamefont {J.~P.}\ \bibnamefont
  {Brantut}}, \bibinfo {author} {\bibfnamefont {J.}~\bibnamefont {Meineke}},
  \bibinfo {author} {\bibfnamefont {D.}~\bibnamefont {Stadler}}, \bibinfo
  {author} {\bibfnamefont {S.}~\bibnamefont {Krinner}}, \ and\ \bibinfo
  {author} {\bibfnamefont {T.}~\bibnamefont {Esslinger}},\ }\href@noop {}
  {\bibfield  {journal} {\bibinfo  {journal} {Science}\ }\textbf {\bibinfo
  {volume} {337}},\ \bibinfo {pages} {1069} (\bibinfo {year}
  {2012})}\BibitemShut {NoStop}%
\bibitem [{\citenamefont {B\"uttiker}(1986)}]{Buettiker_PRL86}%
  \BibitemOpen
  \bibfield  {author} {\bibinfo {author} {\bibfnamefont {M.}~\bibnamefont
  {B\"uttiker}},\ }\href {\doibase 10.1103/PhysRevLett.57.1761} {\bibfield
  {journal} {\bibinfo  {journal} {Phys. Rev. Lett.}\ }\textbf {\bibinfo
  {volume} {57}},\ \bibinfo {pages} {1761} (\bibinfo {year}
  {1986})}\BibitemShut {NoStop}%
\bibitem [{\citenamefont {Blanter}\ and\ \citenamefont
  {M}(2000)}]{Blanter_PR00}%
  \BibitemOpen
  \bibfield  {author} {\bibinfo {author} {\bibfnamefont {Y.~M.}\ \bibnamefont
  {Blanter}}\ and\ \bibinfo {author} {\bibfnamefont {B.}~\bibnamefont {M}},\
  }\href@noop {} {\bibfield  {journal} {\bibinfo  {journal} {Physics Reports}\
  }\textbf {\bibinfo {volume} {336}},\ \bibinfo {pages} {1 } (\bibinfo {year}
  {2000})}\BibitemShut {NoStop}%
\bibitem [{\citenamefont {Boulat}\ and\ \citenamefont
  {Saleur}(2008)}]{Boulat_PRB08}%
  \BibitemOpen
  \bibfield  {author} {\bibinfo {author} {\bibfnamefont {E.}~\bibnamefont
  {Boulat}}\ and\ \bibinfo {author} {\bibfnamefont {H.}~\bibnamefont
  {Saleur}},\ }\href {\doibase 10.1103/PhysRevB.77.033409} {\bibfield
  {journal} {\bibinfo  {journal} {Phys. Rev. B}\ }\textbf {\bibinfo {volume}
  {77}},\ \bibinfo {pages} {033409} (\bibinfo {year} {2008})}\BibitemShut
  {NoStop}%
\bibitem [{\citenamefont {Killi}\ and\ \citenamefont
  {Paramekanti}(2012)}]{Killi_PRA12}%
  \BibitemOpen
  \bibfield  {author} {\bibinfo {author} {\bibfnamefont {M.}~\bibnamefont
  {Killi}}\ and\ \bibinfo {author} {\bibfnamefont {A.}~\bibnamefont
  {Paramekanti}},\ }\href@noop {} {\bibfield  {journal} {\bibinfo  {journal}
  {Phys. Rev. A}\ }\textbf {\bibinfo {volume} {85}},\ \bibinfo {pages} {061606}
  (\bibinfo {year} {2012})}\BibitemShut {NoStop}%
\bibitem [{\citenamefont {Killi}\ \emph {et~al.}(2012)\citenamefont {Killi},
  \citenamefont {Trotzky},\ and\ \citenamefont {Paramekanti}}]{Killi2_PRA12}%
  \BibitemOpen
  \bibfield  {author} {\bibinfo {author} {\bibfnamefont {M.}~\bibnamefont
  {Killi}}, \bibinfo {author} {\bibfnamefont {S.}~\bibnamefont {Trotzky}}, \
  and\ \bibinfo {author} {\bibfnamefont {A.}~\bibnamefont {Paramekanti}},\
  }\href {\doibase 10.1103/PhysRevA.86.063632} {\bibfield  {journal} {\bibinfo
  {journal} {Phys. Rev. A}\ }\textbf {\bibinfo {volume} {86}},\ \bibinfo
  {pages} {063632} (\bibinfo {year} {2012})}\BibitemShut {NoStop}%
\end{thebibliography}%

\end{document}